\documentclass[aps, prl, amsmath, floats,floatfix, twocolumn, superscriptaddress]{revtex4}

\usepackage{amssymb}
\usepackage{amsmath}
\usepackage{verbatim}
\usepackage{mathrsfs}
\usepackage{amsfonts}
\usepackage{latexsym}
\usepackage{epsfig}
\usepackage{color}
\usepackage{graphicx,subfigure}
\usepackage{units}
\usepackage{slashbox}
\usepackage{envmath}
\usepackage{natbib}
\usepackage{hyperref}
\usepackage[utf8]{inputenc}

\begin{document}


\definecolor{orange}{rgb}{0.9,0.45,0}

\newcommand{\re}{\mbox{Re}}
\newcommand{\im}{\mbox{Im}}

\newcommand{\tf}[1]{\textcolor{red}{TF: #1}}
\newcommand{\nsg}[1]{\textcolor{blue}{#1}}
\newcommand{\ch}[1]{\textcolor{green}{CH: #1}}
\newcommand{\fdg}[1]{\textcolor{orange}{FDG: #1}}
\newcommand{\pcd}[1]{\textcolor{magenta}{#1}}
\newcommand{\mz}[1]{\textcolor{cyan}{[\bf MZ: #1]}}

\def\CovDev{D}
\def\Res{{\mathcal R}}
\def\Gammaflat{\hat \Gamma}
\def\metricflat{\hat \gamma}
\def\Dflat{\hat {\mathcal D}}
\def\part_n{\partial_\perp}

\def\Lie{\mathcal{L}}
\def\A{\mathcal{X}}
\def\Aphi{\A_{\phi}}
\def\hAphi{\hat{\A}_{\phi}}
\def\E{\mathcal{E}}
\def\Ham{\mathcal{H}}
\def\M{\mathcal{M}}
\def\R{\mathcal{R}}
\def\p{\partial}

\def\hg{\hat{\gamma}}
\def\hA{\hat{A}}
\def\hD{\hat{D}}
\def\hE{\hat{E}}
\def\hR{\hat{R}}
\def\hcA{\hat{\mathcal{A}}}
\def\hDelt{\hat{\triangle}}

\renewcommand{\t}{\times}

\long\def\symbolfootnote[#1]#2{\begingroup%
\def\thefootnote{\fnsymbol{footnote}}\footnote[#1]{#2}\endgroup}


\title{Non-linear dynamics of spinning bosonic stars: \\ formation and stability} 	
 
\author{N. Sanchis-Gual}
\affiliation{Centro de Astrof\'\i sica e Gravita\c c\~ao - CENTRA, Departamento de F\'\i sica,
Instituto Superior T\'ecnico - IST, Universidade de Lisboa - UL, Avenida
Rovisco Pais 1, 1049-001, Portugal} 

 \author{F. Di Giovanni}
\affiliation{Departamento de
  Astronom\'{\i}a y Astrof\'{\i}sica, Universitat de Val\`encia,
  Dr. Moliner 50, 46100, Burjassot (Val\`encia), Spain}

    \author{M. Zilh\~ao}
\affiliation{Centro de Astrof\'\i sica e Gravita\c c\~ao - CENTRA, Departamento de F\'\i sica,
Instituto Superior T\'ecnico - IST, Universidade de Lisboa - UL, Avenida
Rovisco Pais 1, 1049-001, Portugal}

    \author{C. Herdeiro}
\affiliation{Centro de Astrof\'\i sica e Gravita\c c\~ao - CENTRA, Departamento de F\'\i sica,
Instituto Superior T\'ecnico - IST, Universidade de Lisboa - UL, Avenida
Rovisco Pais 1, 1049-001, Portugal}

 \author{P. Cerd\'a-Dur\'an
}
\affiliation{Departamento de
  Astronom\'{\i}a y Astrof\'{\i}sica, Universitat de Val\`encia,
  Dr. Moliner 50, 46100, Burjassot (Val\`encia), Spain}  

\author{J.~A. Font}
\affiliation{Departamento de
  Astronom\'{\i}a y Astrof\'{\i}sica, Universitat de Val\`encia,
  Dr. Moliner 50, 46100, Burjassot (Val\`encia), Spain}
\affiliation{Observatori Astron\`omic, Universitat de Val\`encia, C/ Catedr\'atico 
  Jos\'e Beltr\'an 2, 46980, Paterna (Val\`encia), Spain}

  \author{E. Radu}
\affiliation{Departamento de F\'{\i}sica da Universidade de Aveiro and 
Centre for Research and Development in Mathematics and Applications (CIDMA), 
Campus de Santiago, 
3810-183 Aveiro, Portugal}


\date{July 2019}


\begin{abstract} 
We perform numerical evolutions of the fully non-linear Einstein--(complex, massive)\,Klein-Gordon and Einstein--(complex)\,Proca systems, to assess the formation and stability of spinning bosonic stars. In the scalar/vector case these are known as boson/Proca stars. Firstly, we consider the formation scenario. Starting with constraint-obeying initial data, describing a dilute, axisymmetric cloud of spinning scalar/Proca field, gravitational collapse towards a spinning star occurs, via gravitational cooling. In the scalar case the formation is transient, even for a non-perturbed initial cloud; a non-axisymmetric instability always develops ejecting all the angular momentum from the scalar star. In the Proca case, by contrast, no instability is observed and the evolutions are compatible with the formation of a spinning Proca star. Secondly, we address the stability of an existing star, a stationary solution of the field equations. In the scalar case, a non-axisymmetric perturbation develops collapsing the star to a spinning black hole. No such instability is found in the Proca case, where the star survives large amplitude perturbations;  moreover, some excited Proca stars decay to, and remain as, fundamental states. Our analysis suggests bosonic stars have different stability properties in the scalar/vector case, which we tentatively relate to their toroidal/spheroidal morphology. A parallelism with instabilities of spinning fluid stars is briefly discussed.
\end{abstract}


\maketitle

\vspace{0.8cm}


{\bf {\em Introduction.}} Recent data from gravitational-wave astronomy~\cite{LIGOScientific:2018mvr}, as well as from electromagnetic VLBI observations near galactic centres~\cite{2018A&A...618L..10G,Akiyama:2019cqa}, support the black hole (BH) \textit{hypothesis}: BHs commonly populate the Cosmos, with masses spanning a range of (at least) 10 orders of magnitude. Yet, the elusiveness of the event horizon, the defining property of a BH, rules out an observational ``proof" of their existence. 
Considering, thus, models of BH mimickers is a valuable tool to understand the uniqueness of BH phenomenology.

Within the landscape of BH mimickers, bosonic stars (BSs) are particularly well motivated.  They arise in simple and physically sound field theoretical models: complex, massive, bosonic fields (scalar~\cite{Kaup:1968zz,Ruffini:1969qy} or vector~\cite{Brito:2015pxa}) minimally coupled to Einstein's gravity. Dynamically, moreover, \textit{static, spherical} BSs, are viable; for some range of parameters, the lowest energy stars - \textit{the fundamental family} (FF) - have a formation mechanism~\cite{Seidel:1993zk,di2018dynamical} and are perturbatively stable~\cite{Gleiser:1988ih,Lee:1988av,Brito:2015pxa,sanchis2017numerical}.  The properties and phenomenology of such static BSs have been considered at length (see $e.g.$ the reviews~\cite{Schunck:2003kk,Liebling:2012fv}), including dynamical situations such as orbiting  binaries, from which gravitational waveforms have been extracted~\cite{palenzuela2007head,palenzuela2008orbital,sanchis2019head}. These studies unveiled a close parallelism in the phenomenology of spherical BSs, regardless of their scalar or vector nature. 

Astrophysically, however, rotation is ubiquitous and should, thus, be included in more realistic models of BSs. Both scalar~\cite{Schunck:1996he,Yoshida:1997qf,Kleihaus:2007vk} and vector~\cite{Brito:2015pxa,Herdeiro:2016tmi} \textit{axisymmetric, spinning} BSs (SBSs) have been constructed and some of their phenomenology has been studied~\cite{Vincent:2015xta,Meliani:2015zta}. Yet, their dynamical and stability properties, a key aspect of their physical viability, have remained essentially unexplored - see the discussion in~\cite{Herdeiro:2014goa}.

In this letter we describe the dynamical properties of SBSs, obtained from fully non-linear numerical simulations of the corresponding Einstein-matter system. We provide evidence that scalar  SBSs in the FF are prone to a non-axisymmetric instability. Thus, such stars are transient states, in a dynamical formation scenario. Assuming an already formed scalar SBS, on the other hand, it collapses into a BH after a non-axially symmetric instability develops.  Vector SBSs (\textit{aka} spinning Proca stars), by contrast, are dynamically robust. In the formation scenario we find no evidence of an instability.  In agreement, for already formed vector SBSs we observe that: $(i)$ even large perturbations are dissipated away; and $(ii)$ some stars in excited families decay to the FF where they remain. This suggests scalar/vector SBSs have different dynamical properties and viability, and their toroidal/spheroidal morphology provides a suggestive interpretation.

{\bf {\em SBSs as stationary solutions.}} 
Scalar and vector BSs, with and without spin, arise as equilibrium states in models with Lagrangian density $\mathcal{L}=R/(16\pi 
G) + \mathcal{L}_{m}$, where $R$ is the Ricci scalar, $G$ is Newton's constant and
\begin{equation}
\mathcal{L}_{m}=-\partial^\alpha \phi \,\partial_\alpha \bar{\phi}-\mu^2 \phi\bar{\phi}  \ ,  \ \mathcal{L}_{m}=-\frac{\mathcal{F}_{\alpha\beta}\bar{\mathcal 
{F}}^{\alpha\beta}}{4}-\frac{\mu^2 } {2}\mathcal{A}_\alpha\bar{\mathcal{A}}^\alpha 
\ ,
\label{matterl}
\end{equation}
describe the scalar and vector cases, respectively. The scalar ($\phi$) and vector ($\mathcal{A}_\alpha$) fields are complex-valued, with conjugation denoted by an overbar, both with mass $\mu$. As usual, $\mathcal{F}=d\mathcal{A}$. Henceforth, units with $G=1=c=\mu$ are used.

\begin{figure}[h!]
\centering
\includegraphics[height=1.35in]{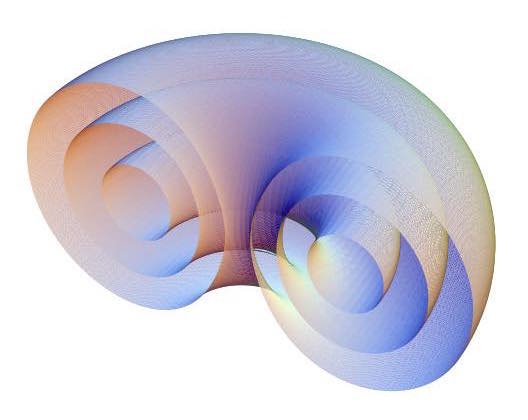}
\includegraphics[height=1.35in]{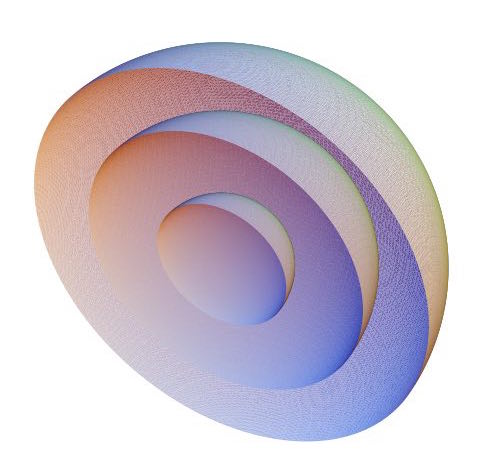}
\caption{Surfaces of constant energy density for illustrative SBSs: (left panel) scalar configuration $2_S$; (right panel) vector configuration $1_P$. The toroidal $vs.$ spheroidal nature is clear.}
\label{fig0}
\end{figure}

Scalar SBSs were first constructed numerically in~\cite{Schunck:1996he,Yoshida:1997qf}, as asymptotically flat, stationary and axisymmetric  solutions of the above Einstein-Klein-Gordon system. They are a ``mass torus" in general relativity  - Fig.~\ref{fig0} (left panel). Scalar SBSs form a discrete set of families of continuous solutions. Each family is labelled by two integers: $m$, the azimuthal winding number and $n$, the node (or overtone) number, see $e.g.$~\cite{Schunck:2003kk,Kleihaus:2007vk,Liebling:2012fv,Grandclement:2014msa,Herdeiro:2015gia}. The FF, which has the lowest energy, has $(m,n)=(1,0)$. Fixing the family, $i.e.$ $(m,n)$,  SBS are characterised by their total mass $M$ and angular momentum $J$. They form a one-dimensional set, often labelled by $M$, and oscillation frequency, $\omega$. The (dynamically) most interesting solutions occur in between the Newtonian limit, $\omega\rightarrow 1$ and $M\rightarrow 0$, and the maximal mass solution. The latter occurs for $\omega\rightarrow \omega_{\rm Mmax}$ ($\simeq 0.775$ for the FF) and the ADM mass becomes highest, $M\rightarrow M_{\rm max}$ ($\simeq 1.315$ for the FF).  In Table~\ref{tab0} we list the properties of two illustrative scalar SBSs used in the simulations below.

\begin{table}[h!]
\begin{center}
\begin{tabular}{|c||cc|cc|ccc|}
\hline
Configuration  & $1_{S}$ & $2_{S}$& $1_{P}$ & $2_{P}$ &  $3_{P}$ & $4_{P}$ & $5_{P}$\\
\hline
\hline
type (S or P)  & S & S  & P & P & P & P & P \\
$n$ & 0 & 0 & 0 & 0  &  1 & 1 & 1 \\
$\omega$ & 0.90 & 0.83  & 0.95 & 0.90 & 0.95 & 0.90 & 0.85 \\
$M$ & 1.119 & 1.281& 0.534 & 0.726  &  1.149 & 1.456 & 1.564  \\
$J$   & 1.153 & 1.338& 0.543 & 0.750 & 1.171 & 1.500 & 1.622\\
\hline
\hline
\end{tabular}
\end{center}
\caption{Physical properties of some illustrative SBSs. The second row identifies if they are scalar ($S$) or vector/Proca ($P$). All solutions have $m=1$ and none has an ergo-region.
\label{tab0}}
\end{table}

Vector SBSs were first reported as excited states $(n=1)$ in~\cite{Brito:2015pxa,Herdeiro:2016tmi}. The FF was considered in~\cite{Herdeiro:2017phl,Herdeiro:2019mbz}. The aforementioned description for scalar SBSs applies, \textit{mutatis mutandis}. An important distinction, however, is that the energy distribution is now spheroidal, rather than toroidal~\cite{Herdeiro:2019mbz} - Fig.~\ref{fig0} (right panel). Moreover, for the FF, $\omega_{\rm Mmax}\simeq 0.562$ and $M_{\rm max}\simeq 1.125$~\cite{Herdeiro:2019mbz}. For the excited family with $(m,n)=(1,1)$, $\omega_{\rm Mmax}\simeq 0.839$ and $M_{\rm max}\simeq 1.568$~\cite{Brito:2015pxa}. In Table~\ref{tab0} we list the properties of two [three] representative vector SBSs, in the FF [$(m,n)=(1,1)$ family].

{\bf {\em Dynamical formation of SBSs.}}
In the spherical case,  numerical simulations established that both scalar~\cite{Seidel:1993zk} and vector~\cite{di2018dynamical} BSs form dynamically, from a spherical ``cloud" of dilute scalar or vector field. The cloud collapses due to its self-gravity. The ejection of energetic scalar or vector ``particles", dubbed gravitational cooling, allows the formation of a compact object. 

For studying the formation of SBSs, with $m=1$, the Hamiltonian, momentum and (in the vector case) Gauss constraint are solved by appropriately choosing a Gaussian radial dependence for the key variables, together with a non-spherical profile (see Appendix A). For the scalar case the ``matter" initial data is:
 \begin{equation}\label{Phi}
 \phi(t,r,\theta,\varphi)= A\, r\, e^{-\frac{r^2}{\sigma^2}} \sin{\theta}\,e^{i(\varphi-\omega t)} \ ,
\end{equation}
where $A$, $\sigma$ are constants and $e^{-i\omega t}$ is the harmonic dependence. Besides this unperturbed initial data, we also evolve perturbed initial data of two types: replacing in~\eqref{Phi} $e^{i\varphi} \rightarrow e^{i\varphi}(1+A_1\cos(2\varphi))$; or, alternatively, replacing $e^{i\varphi} \rightarrow  e^{i\varphi}+A_2e^{2i\varphi}$. $A_1,A_2$ are the amplitudes of the perturbations. 

Fully non-linear numerical evolutions of the Einstein-matter system using this initial data were carried out with the \textsc{Einstein Toolkit}~\cite{EinsteinToolkit:web,loffler2012f} - see Appendix B. Two choices of $A$ were considered, both of which yield global data for the scalar cloud $(M_{\rm{Sc}},J_{\rm{Sc}})$ close to that of equilibrium scalar SBS solutions. The first/second choices give $M_{\rm{Sc}}^{(1)}\sim0.46\sim J_{\rm{Sc}}^{(1)}$, and $M_{\rm{Sc}}^{(2)}\sim0.89\sim J_{\rm{Sc}}^{(2)}$, respectively. We have run simulations with both perturbed and unperturbed initial data, with $A_1= 0, 0.001,0.01,0.05$ and $A_2= 0,0.05$. Typically, $\sigma=40$.

All evolutions show the emergence of a non-axisymmetric instability. The time at which the instability kicks in depends on the type and amplitude of the perturbation; but even the lowest mass unperturbed model ($M_{\rm{Sc}}^{(1)}$) exhibits non-axisymmetric features at a sufficiently long time scale ($t\sim {\cal O}(10^4)$). The instability generically triggers a larger ejection of angular momentum than mass, reshaping the toroidal energy distribution into a spherical one. This suggests that the asymptotic end state of the cloud evolution is either a spherical (non-spinning) scalar BS or even, merely, ejected debris carrying all angular momentum and energy. 

\begin{figure}[h!]
\begin{tabular}{ p{0.5\linewidth}  p{0.5\linewidth} }
\centering Scalar SBS & \centering Vector SBS
\end{tabular}
\\
\includegraphics[width=0.24\linewidth]{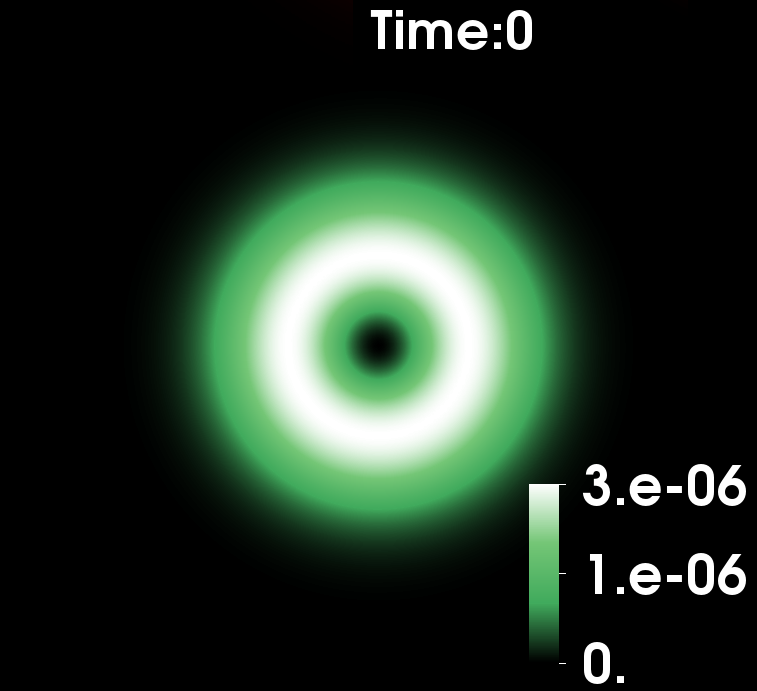}\hspace{-0.01\linewidth}
\includegraphics[width=0.24\linewidth]{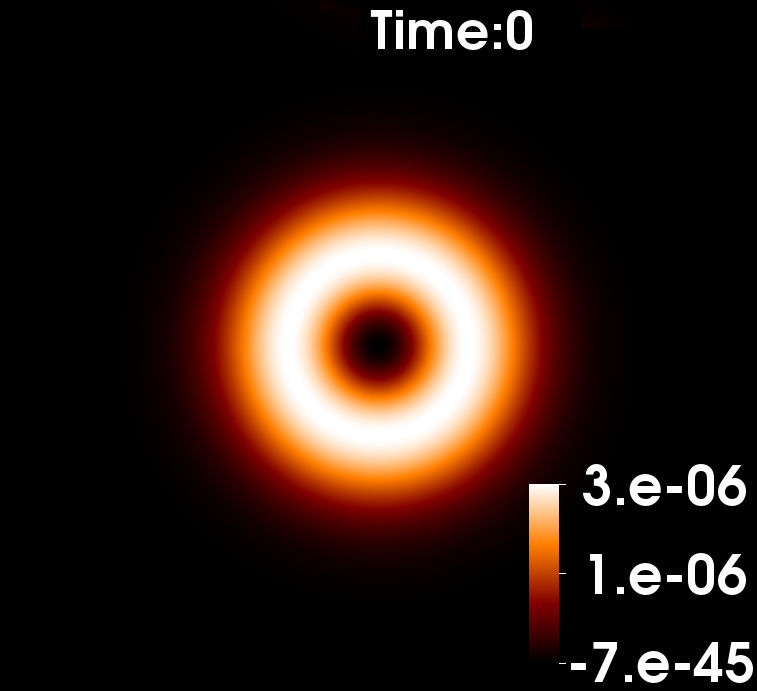}\hspace{0.02\linewidth}
\includegraphics[width=0.24\linewidth]{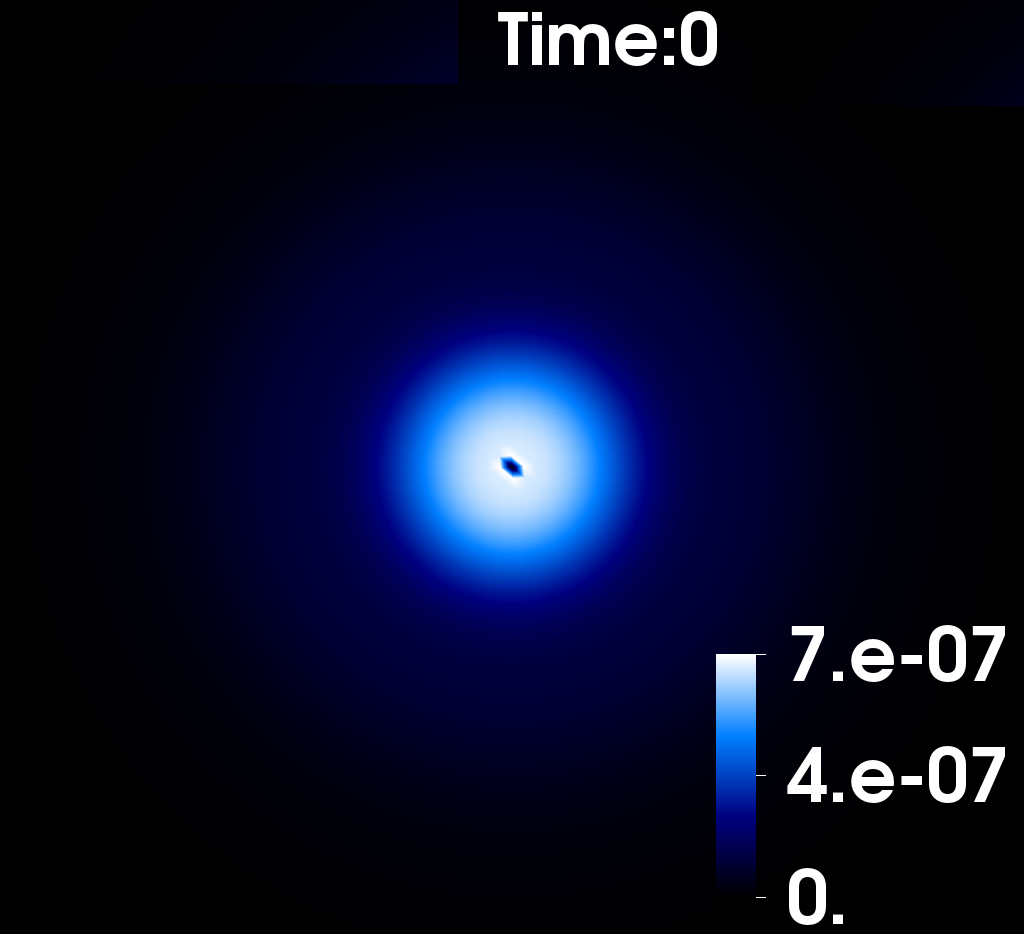}\hspace{-0.01\linewidth}
\includegraphics[width=0.24\linewidth]{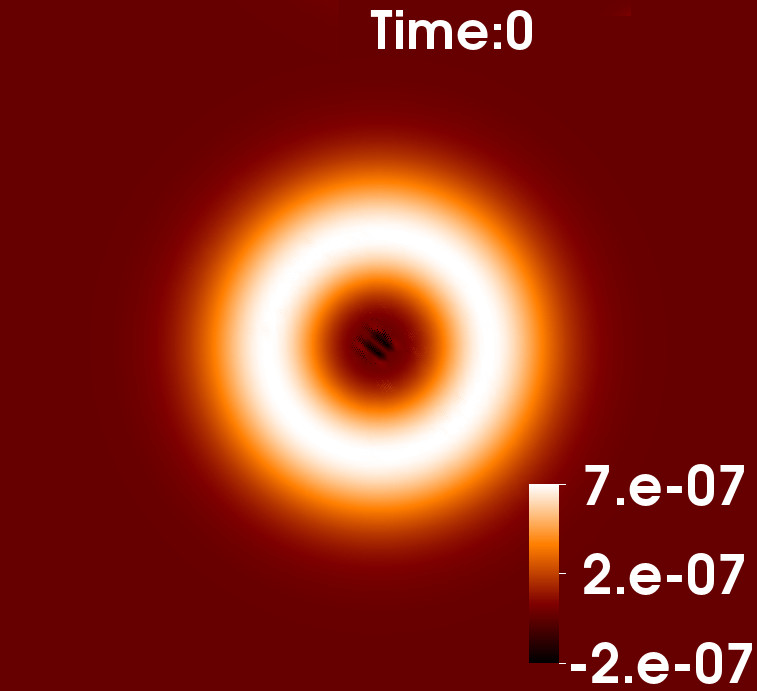}\\
\includegraphics[width=0.24\linewidth]{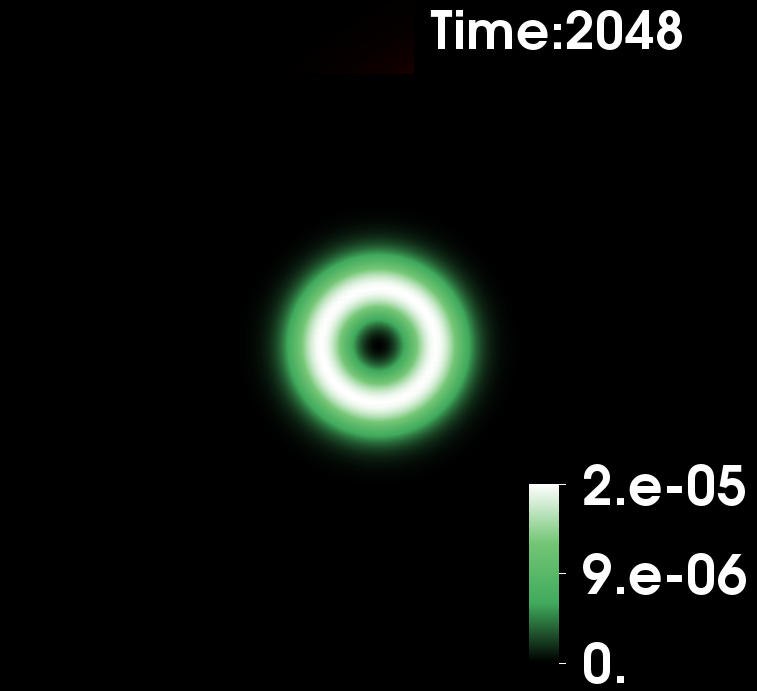}\hspace{-0.01\linewidth}
\includegraphics[width=0.24\linewidth]{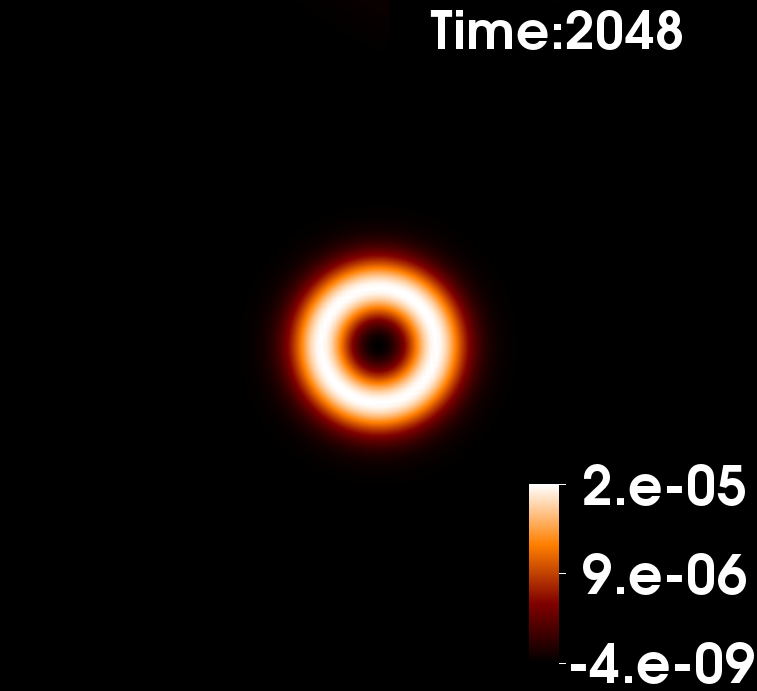}\hspace{0.02\linewidth}
\includegraphics[width=0.24\linewidth]{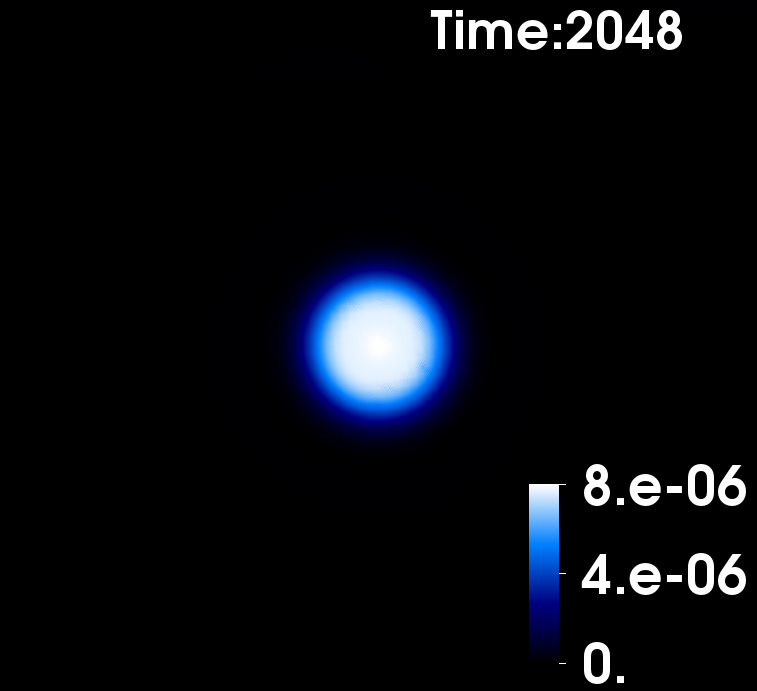}\hspace{-0.01\linewidth}
\includegraphics[width=0.24\linewidth]{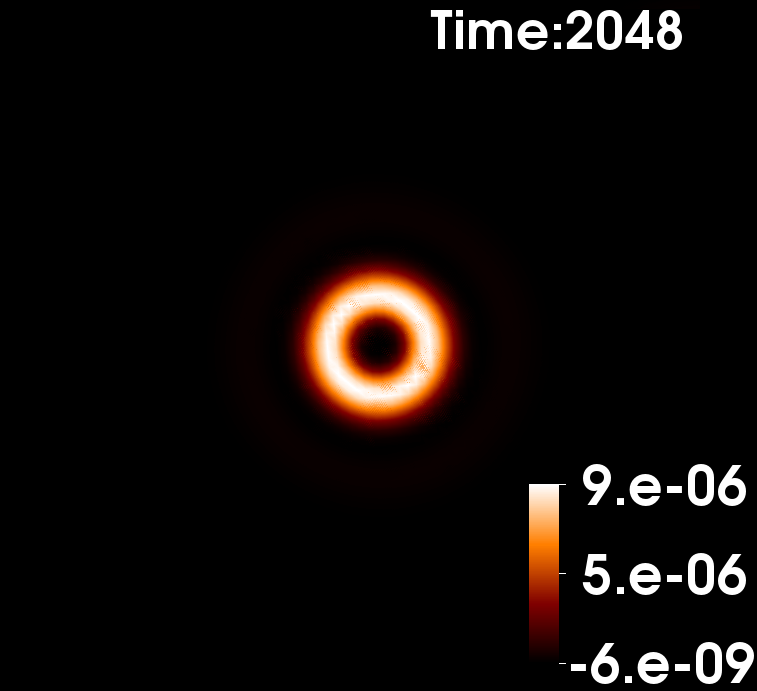}\\
\includegraphics[width=0.24\linewidth]{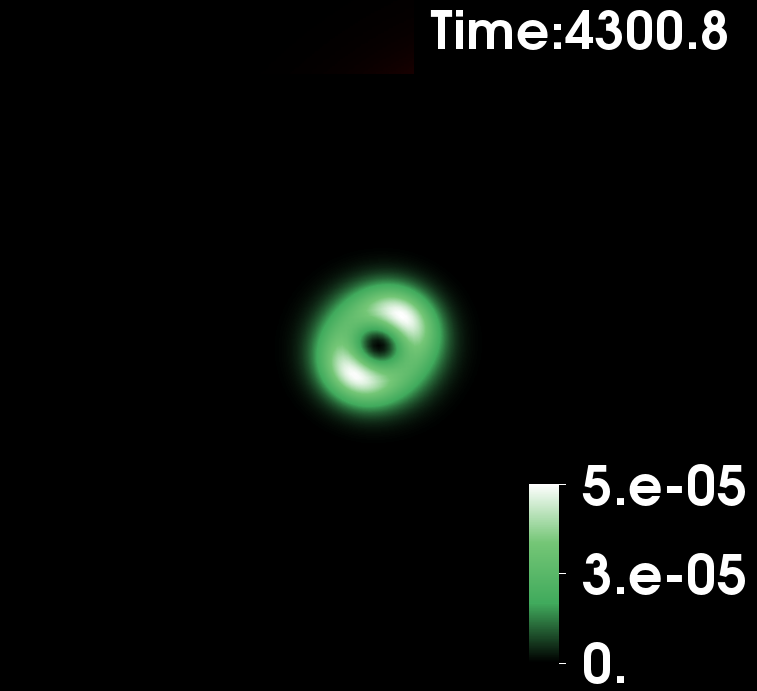}\hspace{-0.01\linewidth}
\includegraphics[width=0.24\linewidth]{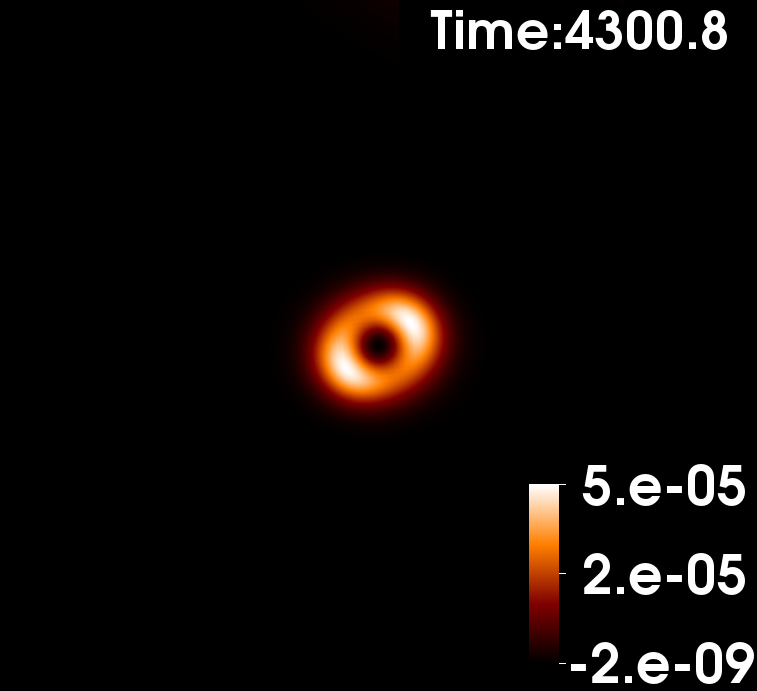}\hspace{0.02\linewidth}
\includegraphics[width=0.24\linewidth]{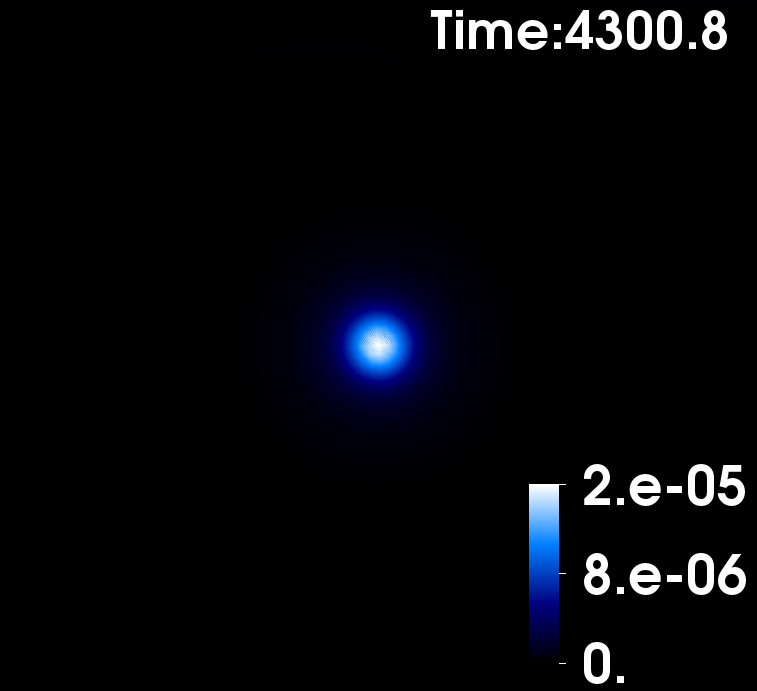}\hspace{-0.01\linewidth}
\includegraphics[width=0.24\linewidth]{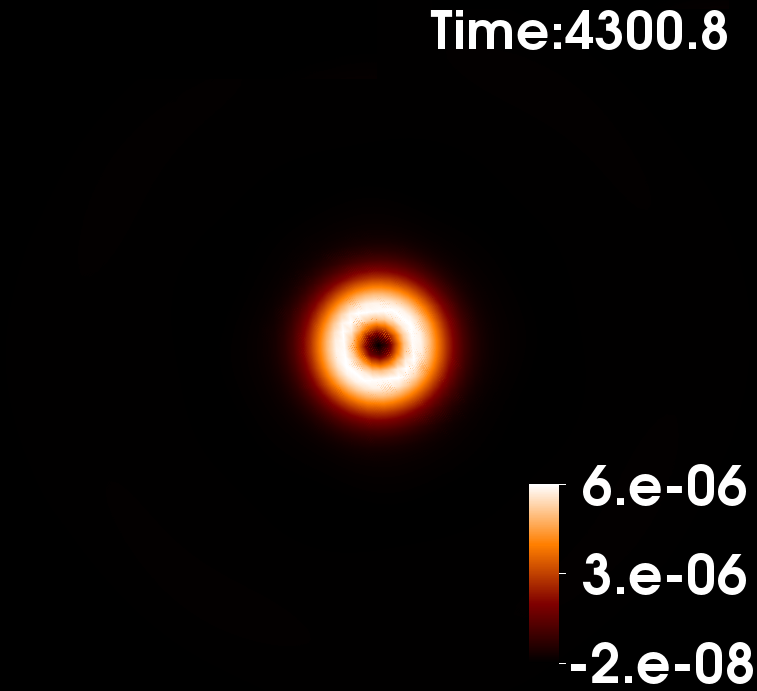}\\
\includegraphics[width=0.24\linewidth]{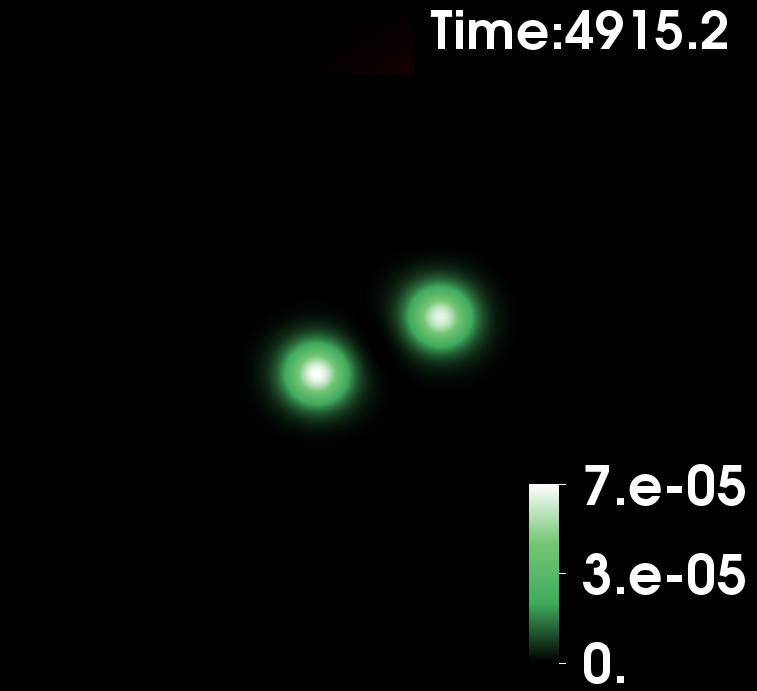}\hspace{-0.01\linewidth}
\includegraphics[width=0.24\linewidth]{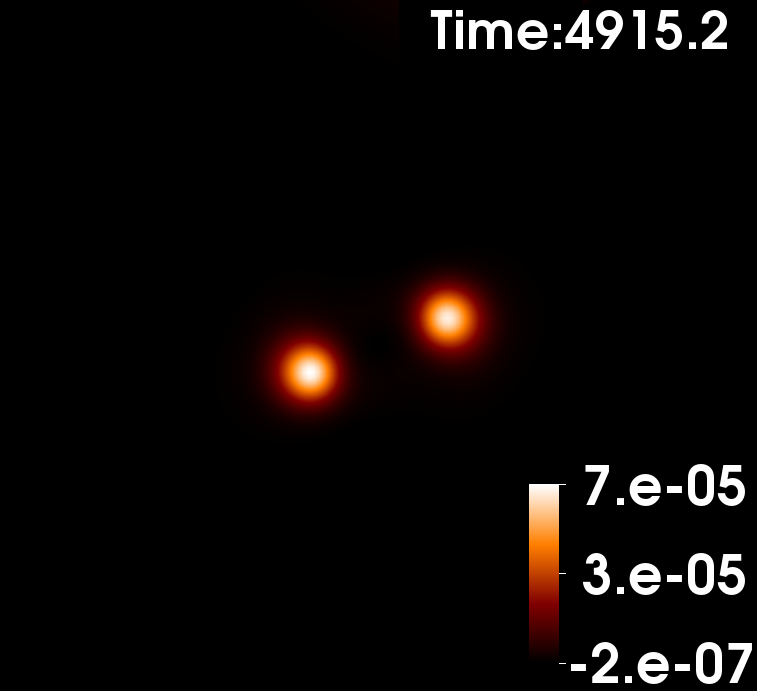}\hspace{0.02\linewidth}
\includegraphics[width=0.24\linewidth]{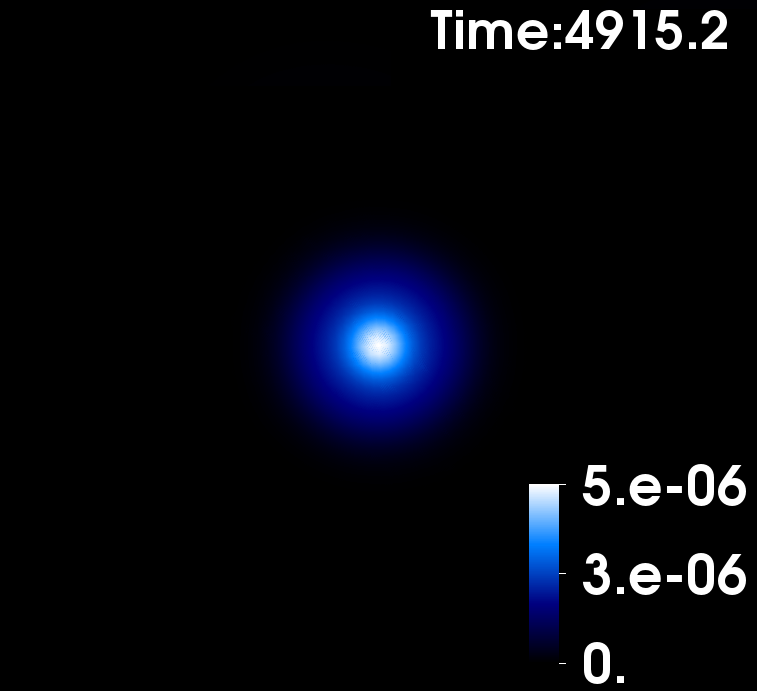}\hspace{-0.01\linewidth}
\includegraphics[width=0.24\linewidth]{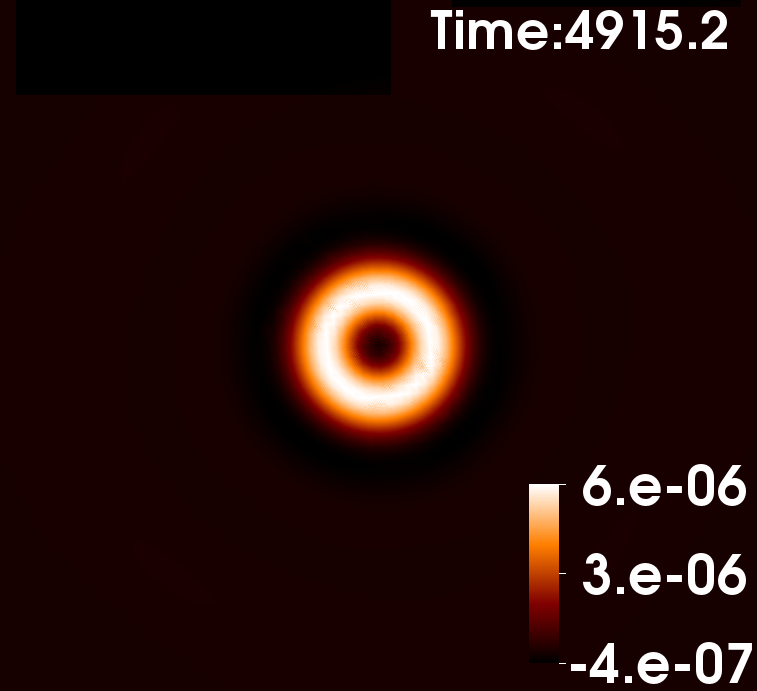}\\
\includegraphics[width=0.24\linewidth]{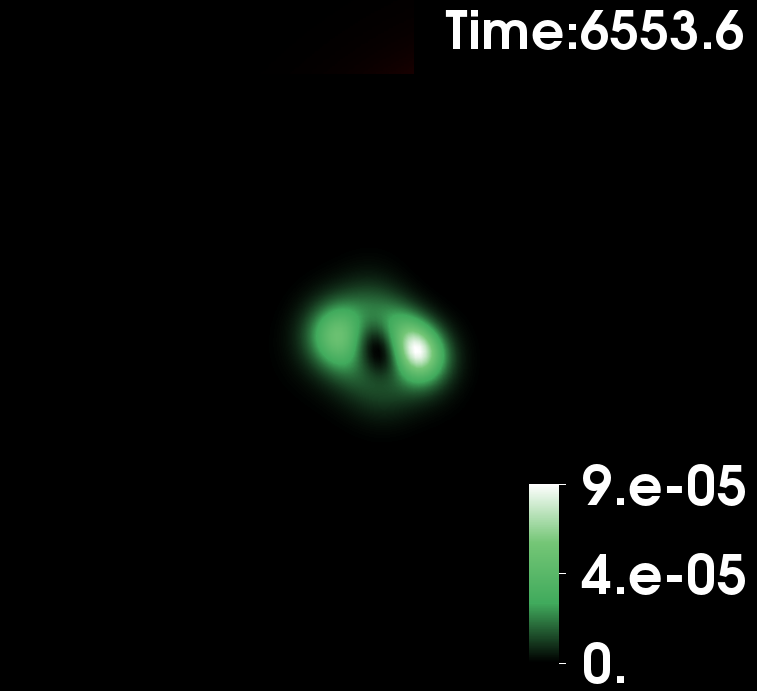}\hspace{-0.01\linewidth}
\includegraphics[width=0.24\linewidth]{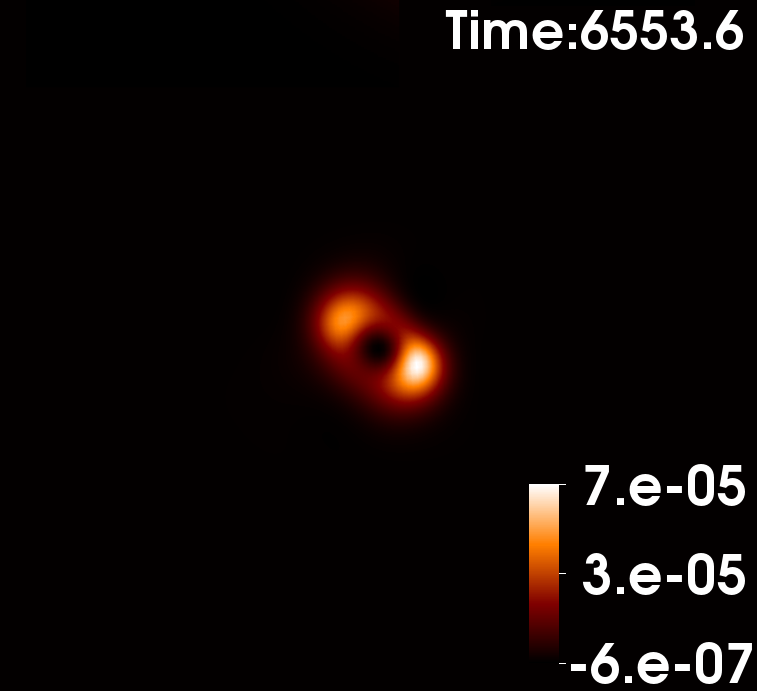}\hspace{0.02\linewidth}
\includegraphics[width=0.24\linewidth]{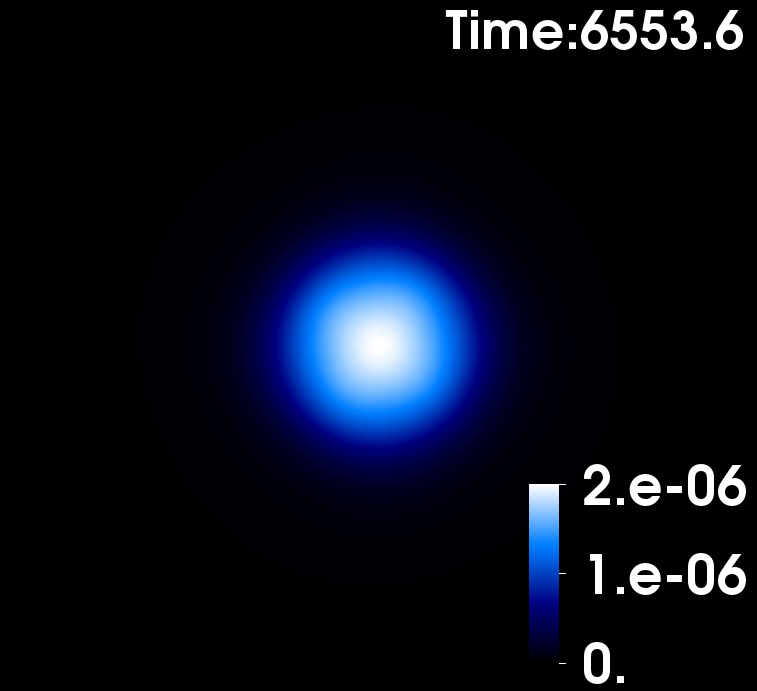}\hspace{-0.01\linewidth}
\includegraphics[width=0.24\linewidth]{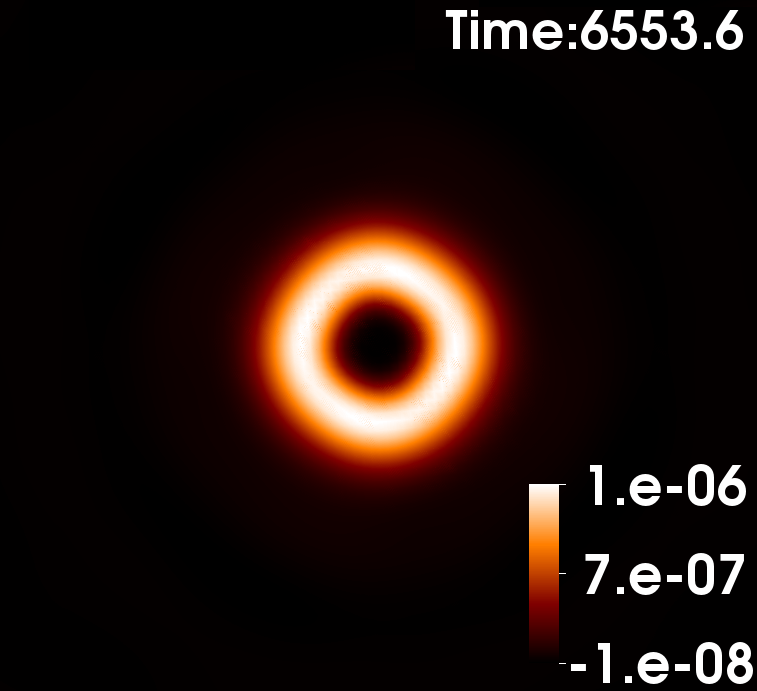}\\
\includegraphics[width=0.24\linewidth]{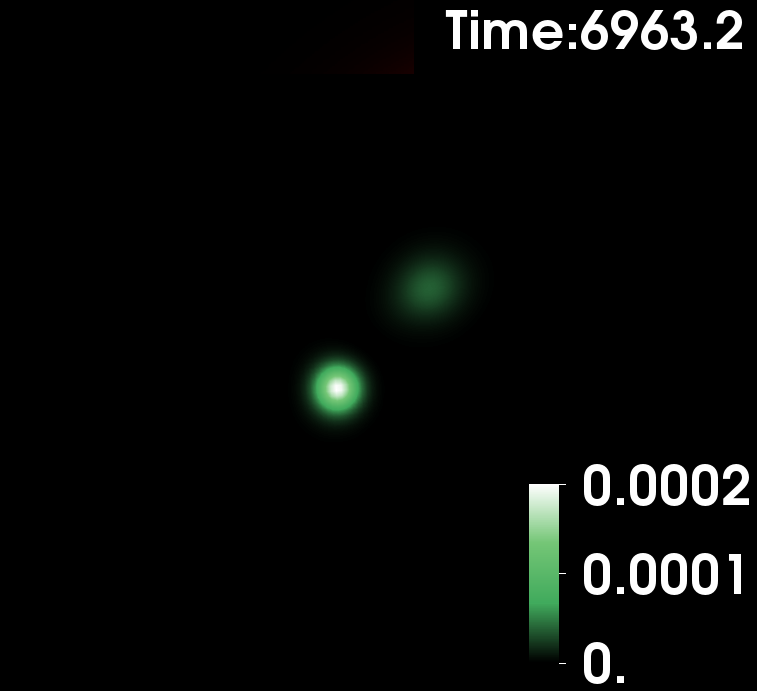}\hspace{-0.01\linewidth}
\includegraphics[width=0.24\linewidth]{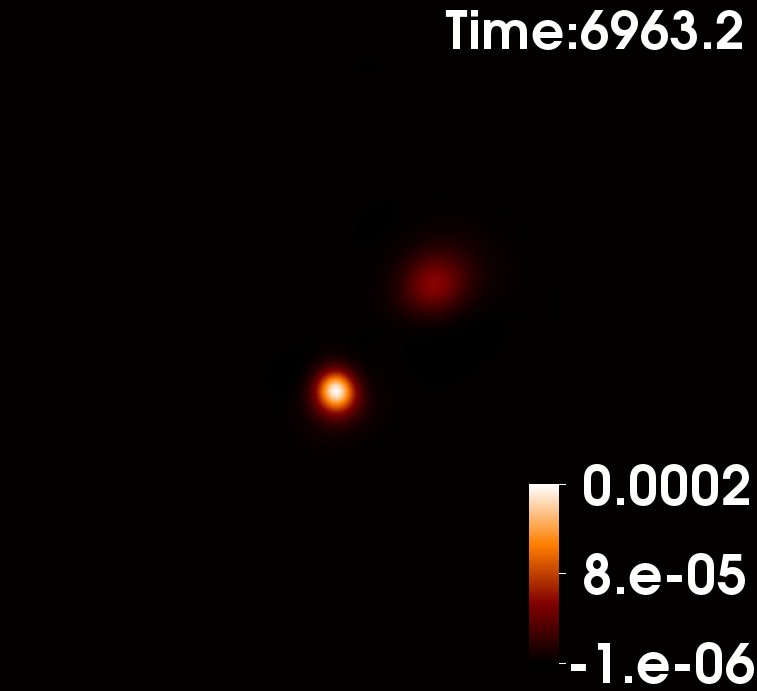}\hspace{0.02\linewidth}
\includegraphics[width=0.24\linewidth]{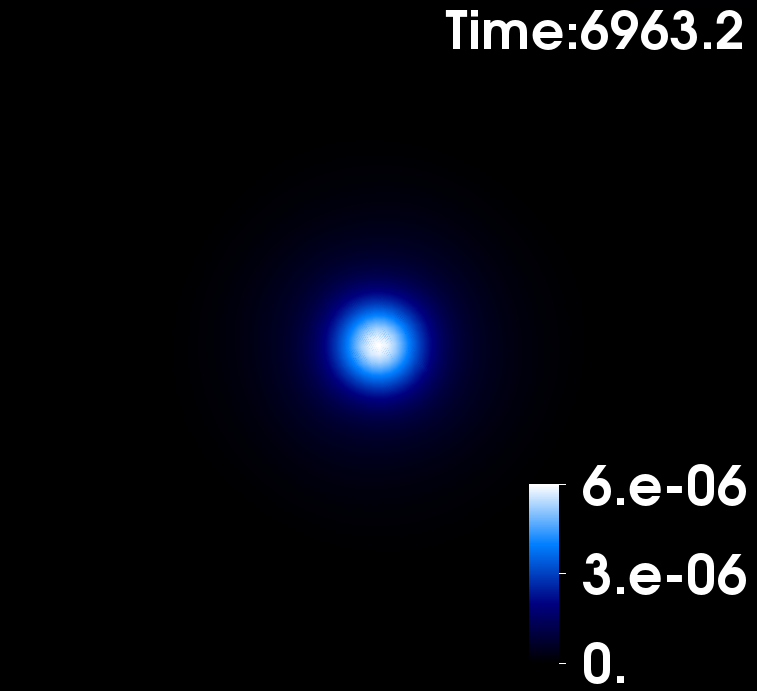}\hspace{-0.01\linewidth}
\includegraphics[width=0.24\linewidth]{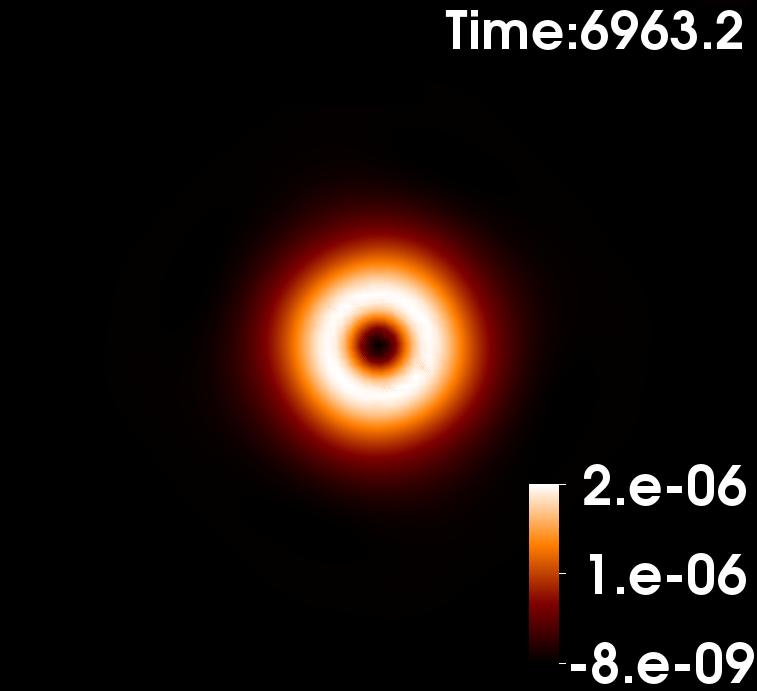}\\
\includegraphics[width=0.24\linewidth]{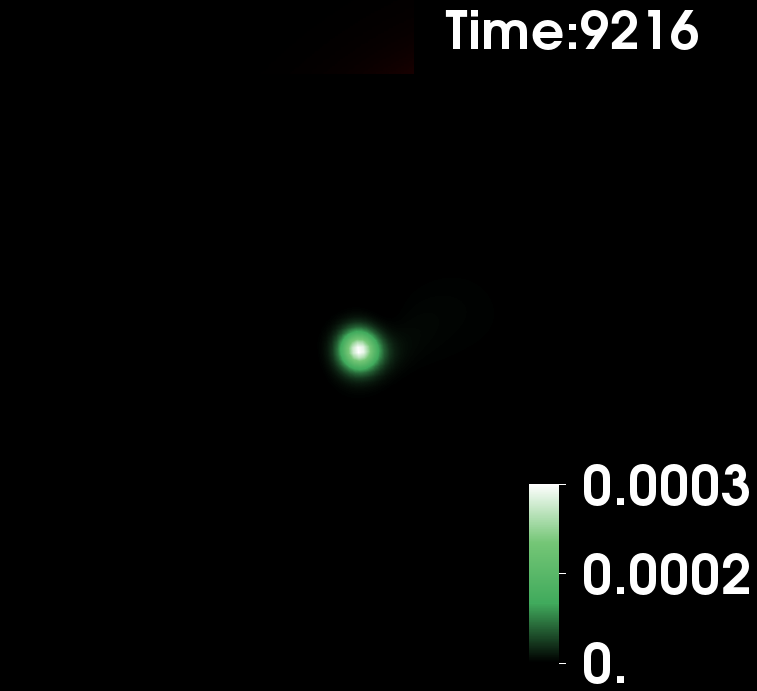}\hspace{-0.01\linewidth}
\includegraphics[width=0.24\linewidth]{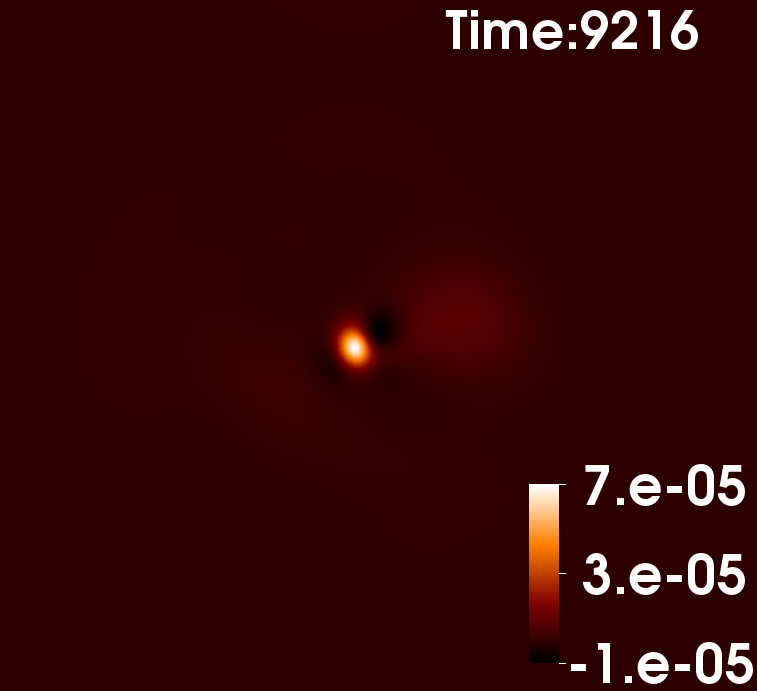}\hspace{0.02\linewidth}
\includegraphics[width=0.24\linewidth]{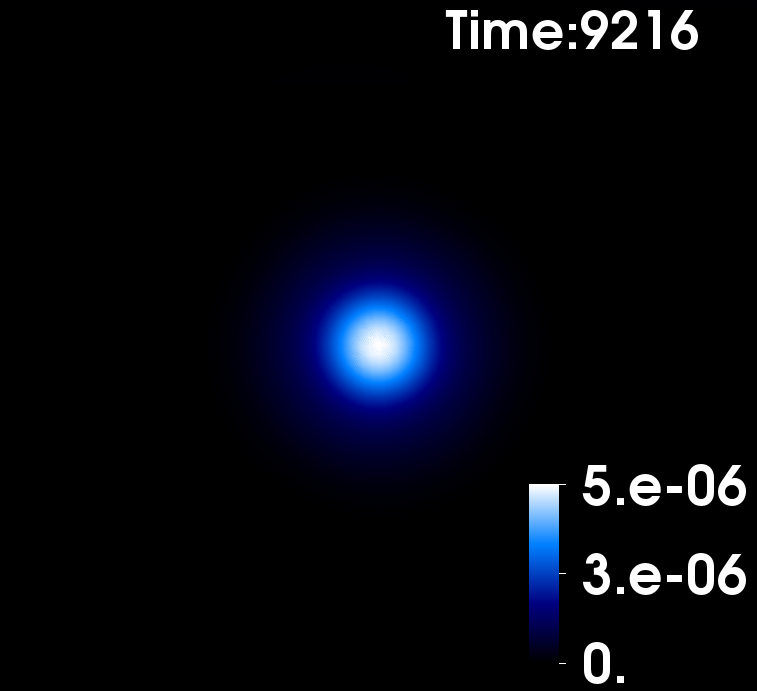}\hspace{-0.01\linewidth}
\includegraphics[width=0.24\linewidth]{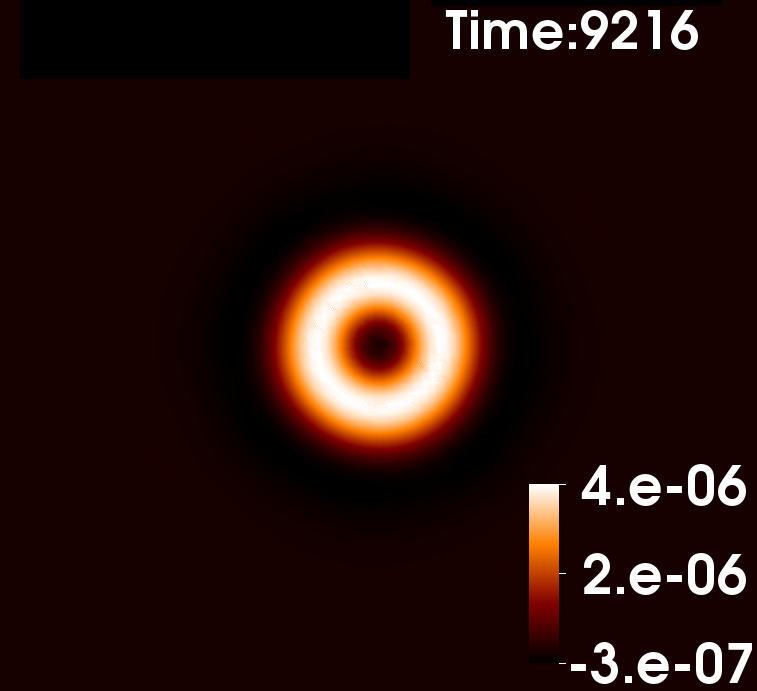}\\
\caption{Time evolution of an equatorial cut of $\rho_E$ (blue/green) and $\rho_J$ (orange) in the formation scenario of a scalar (left side) or a vector (right side) SBS.
}
\label{fig2}
\end{figure}

As an illustration, Fig.~\ref{fig2} exhibits snapshots of the equatorial plane evolution of the energy density, $\rho_E$ (left panels) and angular momentum density, $\rho_J$ (middle-left panels), for the unperturbed scalar initial data with mass $M_{\rm{Sc}}^{(2)}$~\footnote{The energy and angular momentum densities are the ones defined from the Komar integrals, as in~\cite{Cunha:2017wao,sanchis2019head}.}. Initially, the collapse preserves axial symmetry. 
Around $t\sim 4000$, however, the non-axisymmetric instability is visible, producing a fragmentation event: the star splits into a roughly symmetric orbiting binary. The binary is, nonetheless,  bound and recollapses to a deformed spinning star, around $t\sim6500$. This star breaks into two asymmetric pieces, which again recollapse into a spheroidal star with angular momentum. Around $t\sim 10000$, this residual, still evolving, star has $(M, J)$= (0.49, 0.16), evaluated up to $r=30$, and an oscillation frequency $\omega\sim0.96$. For this $\omega$, the FF static scalar BS has $(M, J)= (0.45, 0)$. Thus, this (or a neighbour) static scalar BS appears to be asymptotically approached, after the remaining $J$ is shed away.

Now consider the formation of a vector SBS. The construction of initial data is more complex due to the Gauss constraint~\cite{Zilhao:2015tya,di2018dynamical}. After a 3+1 splitting of $\mathcal{A}_\mu$, the key variables are the scalar and 3-vector potentials together with the electric field. The first of these admits a solution almost identical to~\eqref{Phi}, but the others are more involved - Appendix A.  These initial data can again be perturbed. We have considered a perturbation analogous to the first type considered in the scalar case;  the perturbation amplitudes studied were $A_1=A_2=0,0.05$. Initial data describing a Proca cloud with three different values of global data were used: $M_{\rm{Pc}}^{(1)}\sim 0.46 \sim J_{\rm{Pc}}^{(1)}$, $M_{\rm{Pc}}^{(2)}\sim 0.56 \sim J_{\rm{Pc}}^{(2)}$ and $M_{\rm{Pc}}^{(3)}\sim 0.77 \sim J_{\rm{Pc}}^{(3)}$.

The  unperturbed models evolutions are instability-free during the simulations, lasting up to $t\sim 10^4$.  This is illustrated by the 3$^{\rm rd}$ and 4$^{\rm th}$ columns in Fig.~\ref{fig2} which show snapshots of the time evolution of the unperturbed Proca cloud $M_{\rm{Pc}}^{(2)}$. The gravitational collapse ejects part of the mass and  angular momentum, which shows the gravitational cooling mechanism at play. At $t\sim 10^4$ the star has $(M, J)$= (0.35, 0.38), evaluated up to $r=60$, and $\omega\sim0.985$. For this $\omega$, the FF vector SBS has $(M, J)= (0.301, 0.303)$. Thus, this (or a neighbour) vector SBS appears to be asymptotically approached. For the perturbed initial Proca clouds, on the other hand, the energy density oscillates strongly. Nonetheless, no sudden loss of angular momentum is observed, which suggests the endpoint is still a vector SBS.

{\bf {\em Evolution of equilibrium SBSs.}} 
The dichotomy observed in the formation scenario can be further assessed by considering the dynamics of SBSs obtained as equilibrium solutions of the corresponding Einstein-matter system. A perturbative stability analysis of these SBSs, such as the ones in~\cite{Gleiser:1988ih,Lee:1988av,Brito:2015pxa} for the spherical case, seems challenging. Thus, we resort to non-linear numerical evolutions of the Einstein-matter system, analogue to the ones in the formation scenario, but now starting with the equilibrium solutions as initial data. This generalises the evolutions in~\cite{Cunha:2017wao} for non-spinning BSs.

\begin{figure}[h!]
\centering
\includegraphics[width=0.24\linewidth]{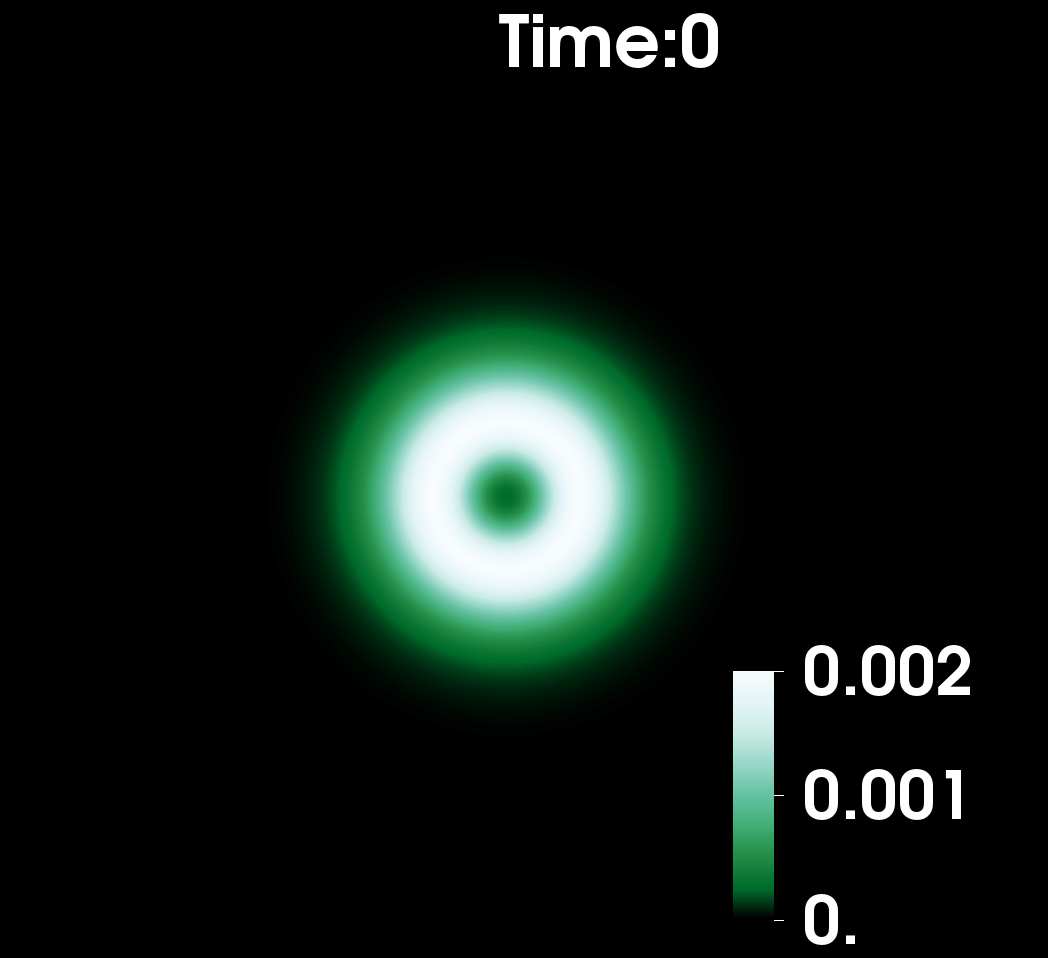}\hspace{-0.01\linewidth}
\includegraphics[width=0.24\linewidth]{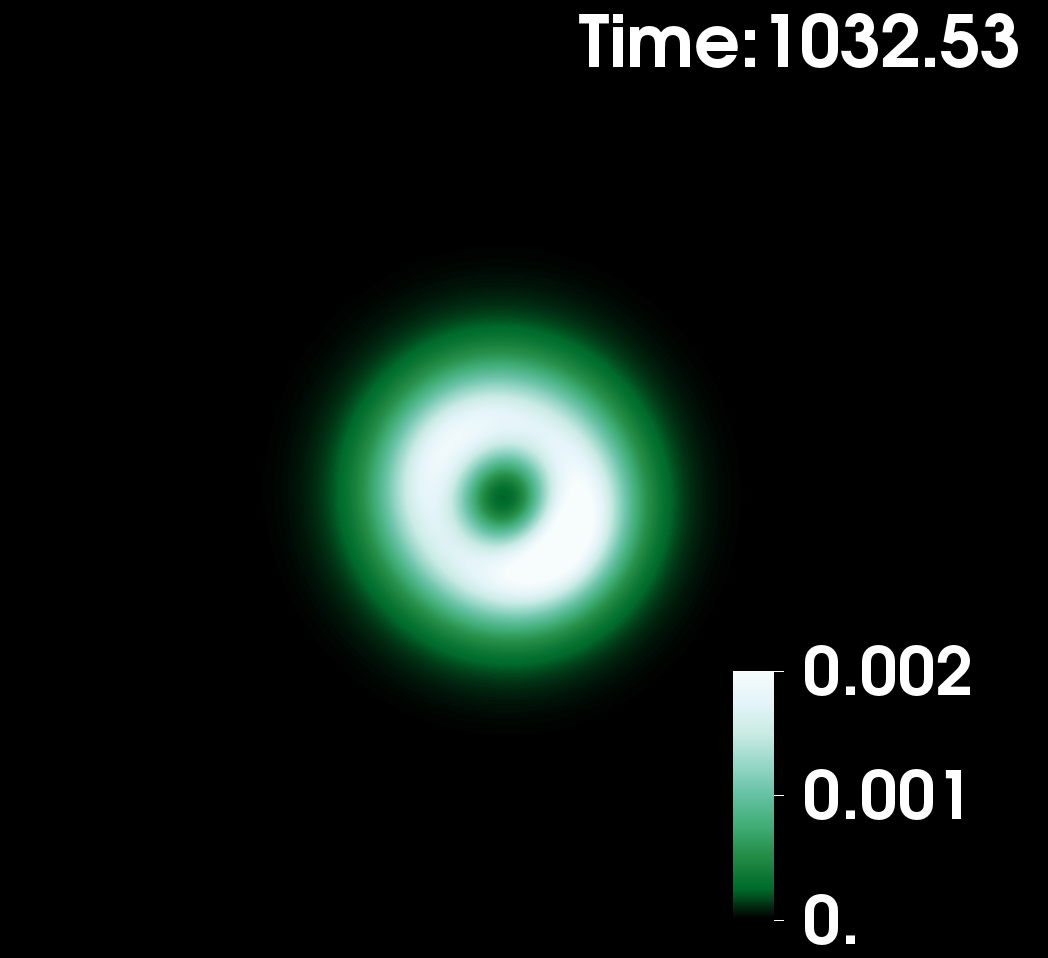}\hspace{-0.01\linewidth}
\includegraphics[width=0.24\linewidth]{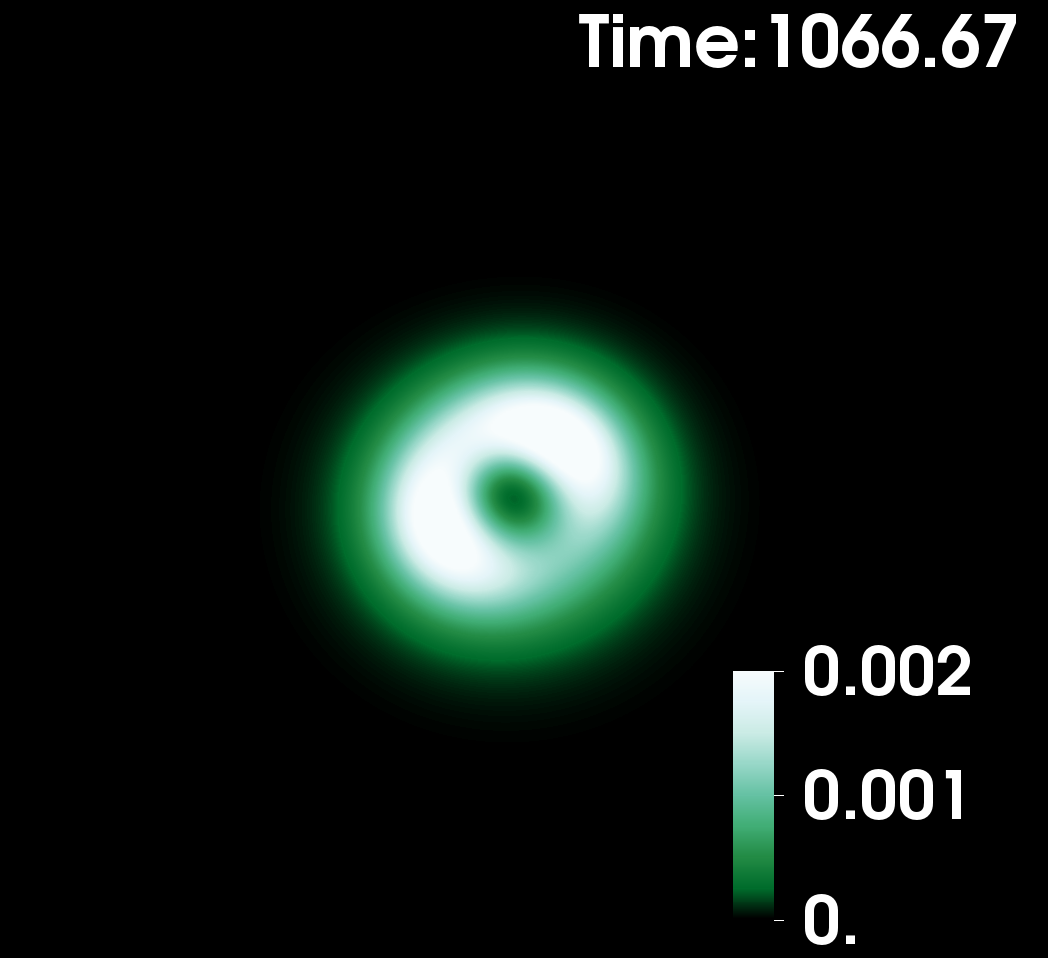}\hspace{-0.01\linewidth}
\includegraphics[width=0.24\linewidth]{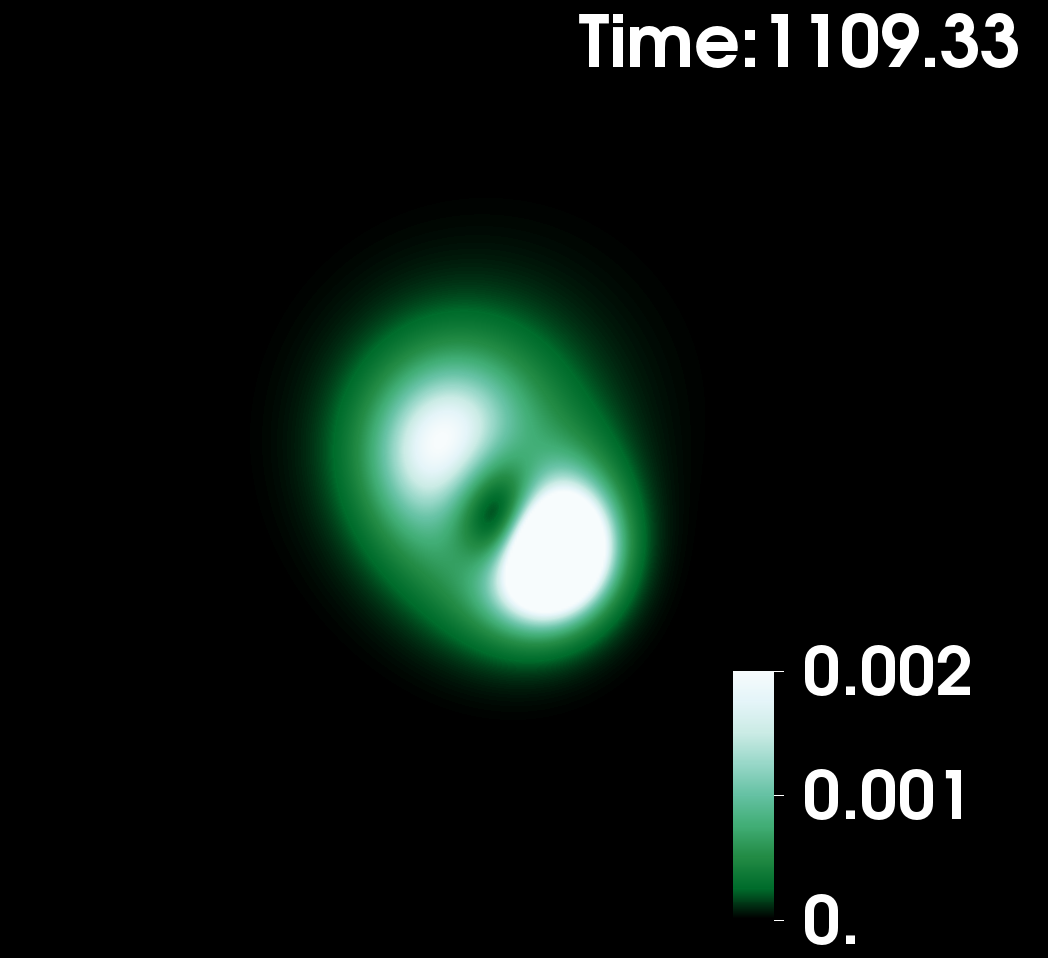}\hspace{0.02\linewidth}\\
\includegraphics[width=0.24\linewidth]{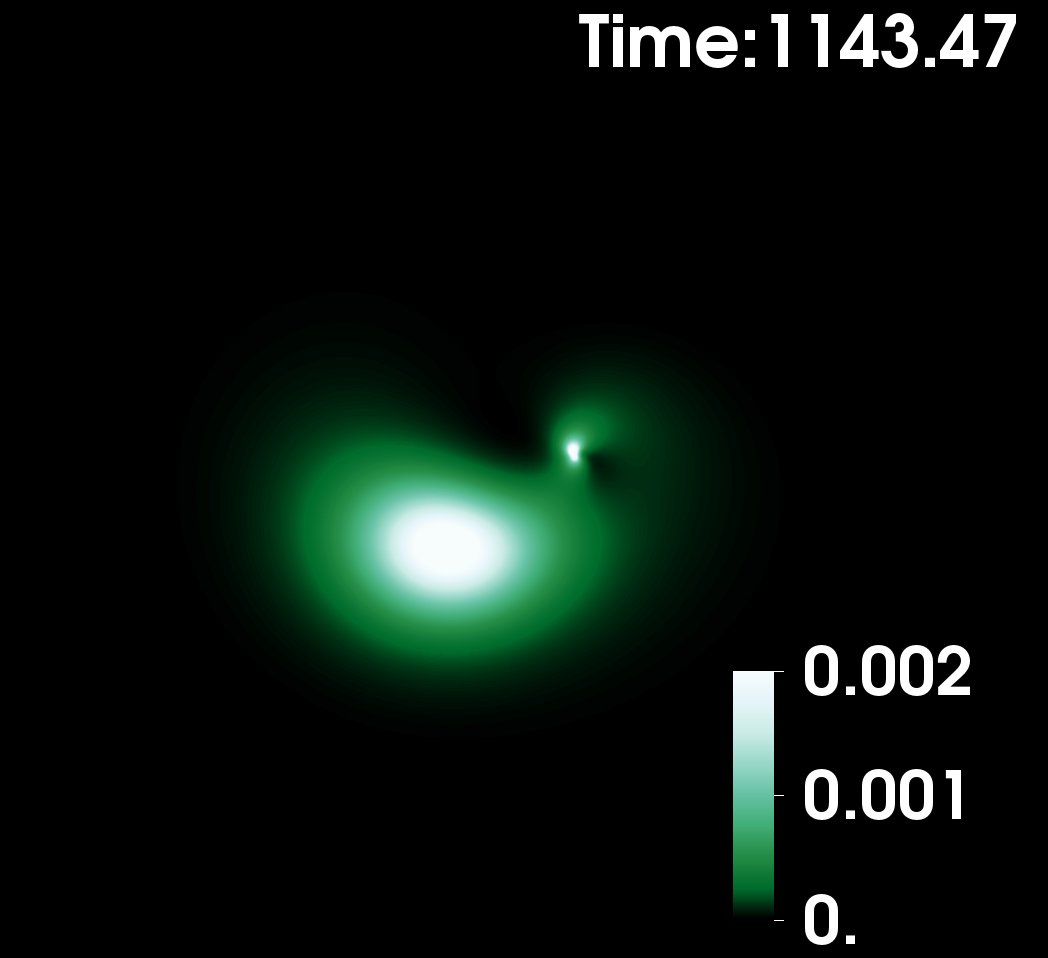}\hspace{-0.01\linewidth}
\includegraphics[width=0.24\linewidth]{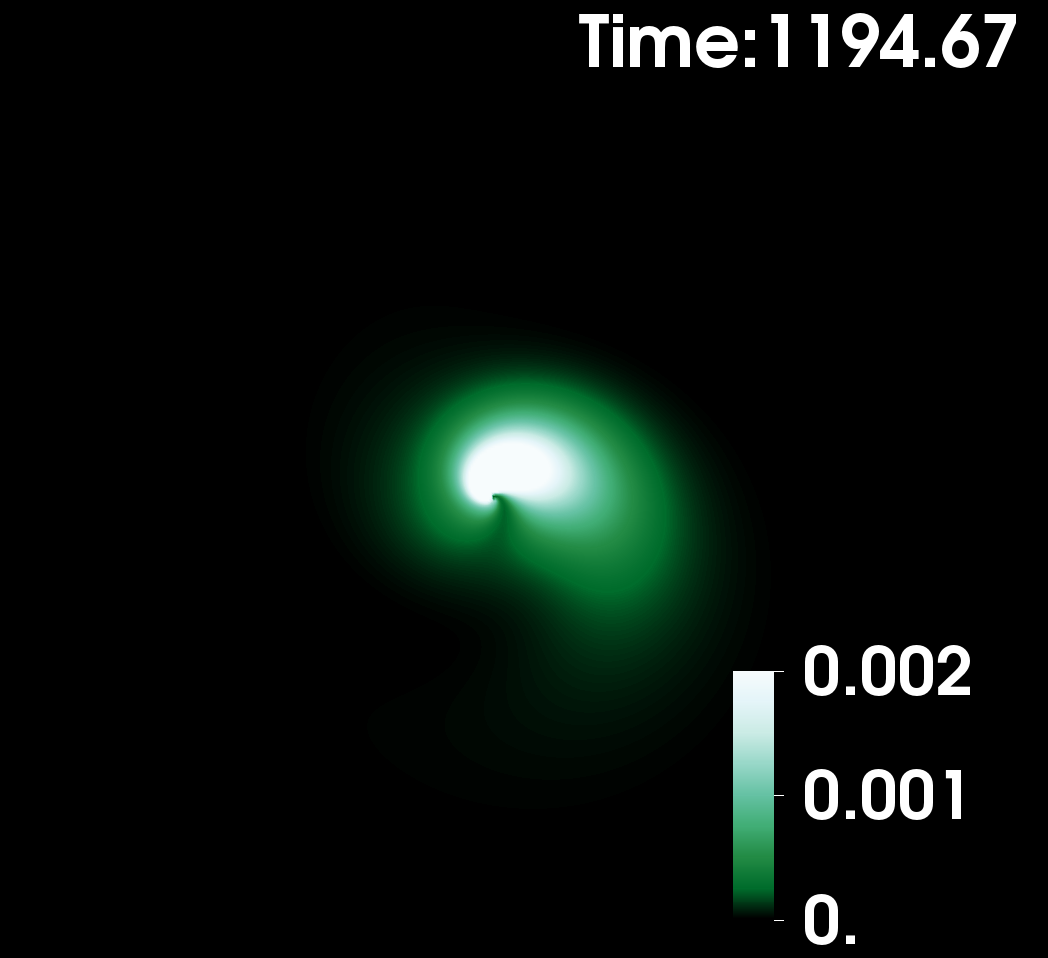}\hspace{-0.01\linewidth}
\includegraphics[width=0.24\linewidth]{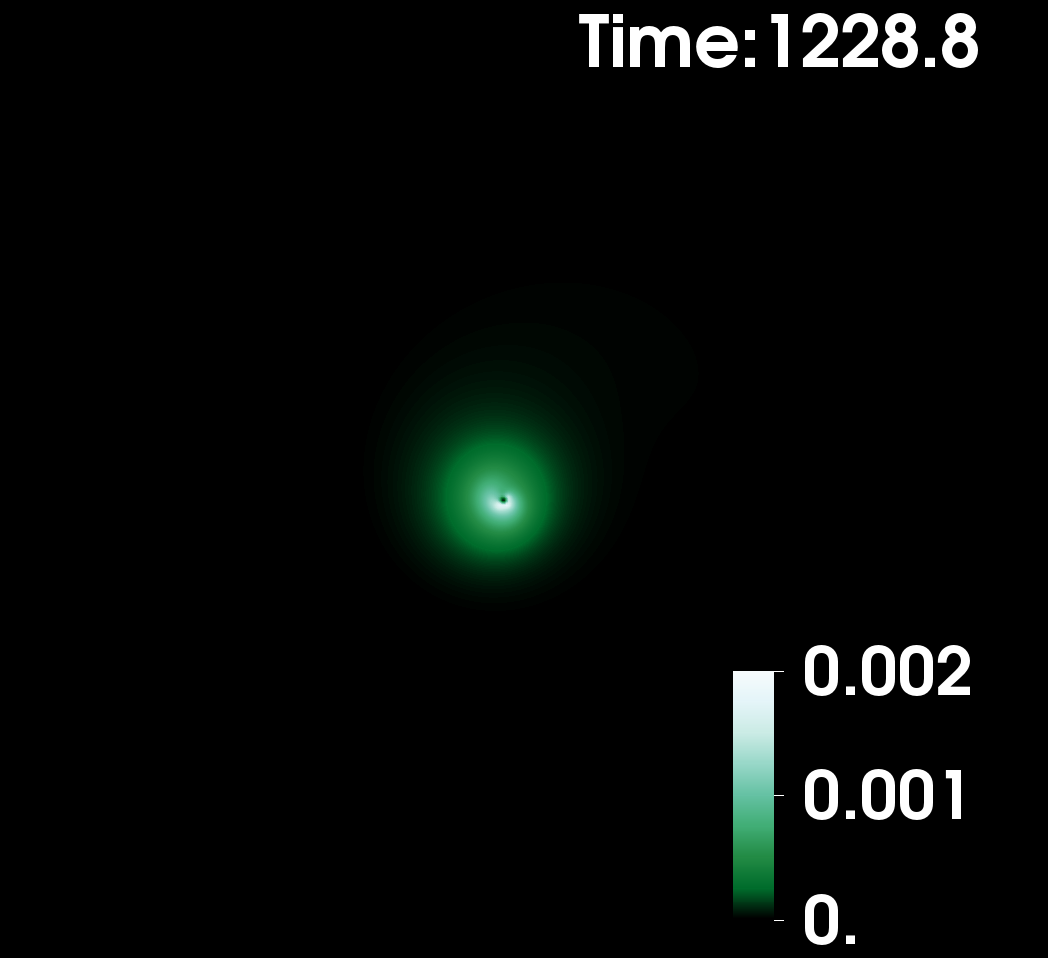}\hspace{-0.01\linewidth}
\includegraphics[width=0.24\linewidth]{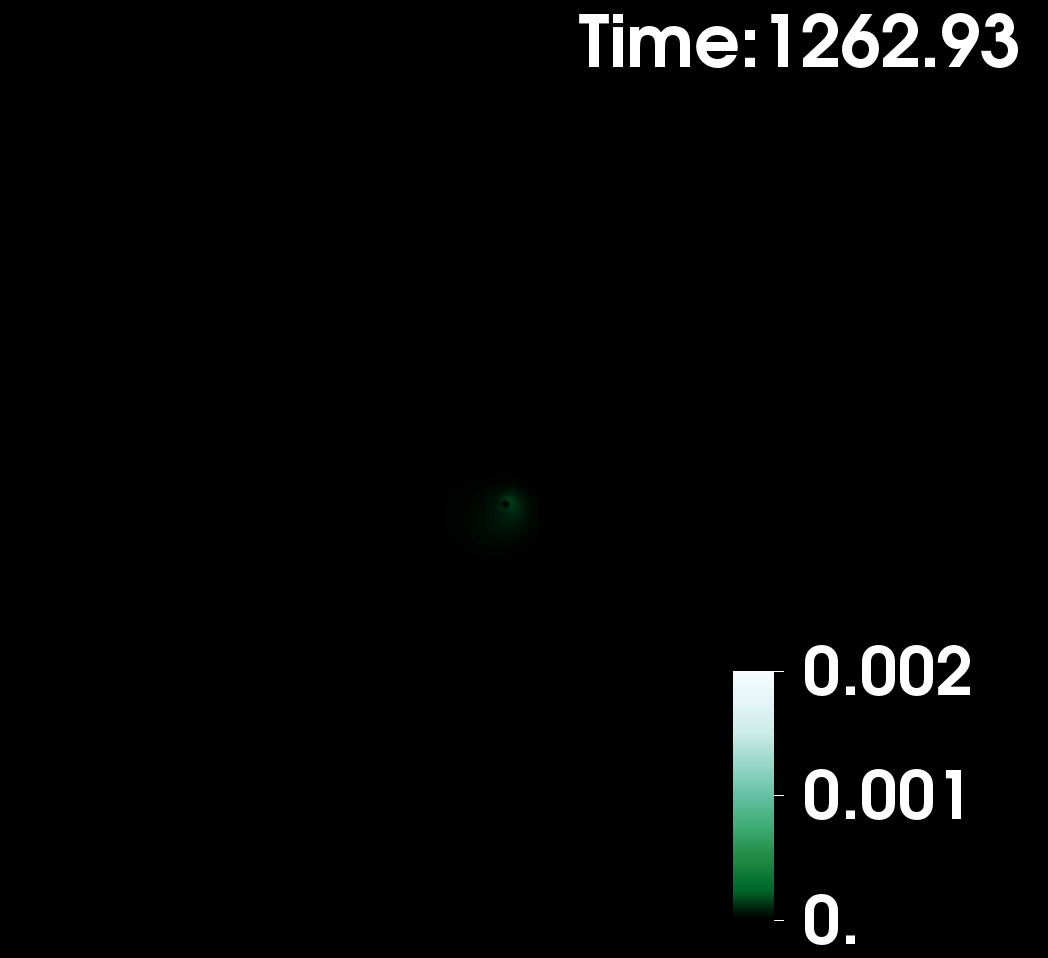}\hspace{0.02\linewidth}\\
\includegraphics[height=2.4in]{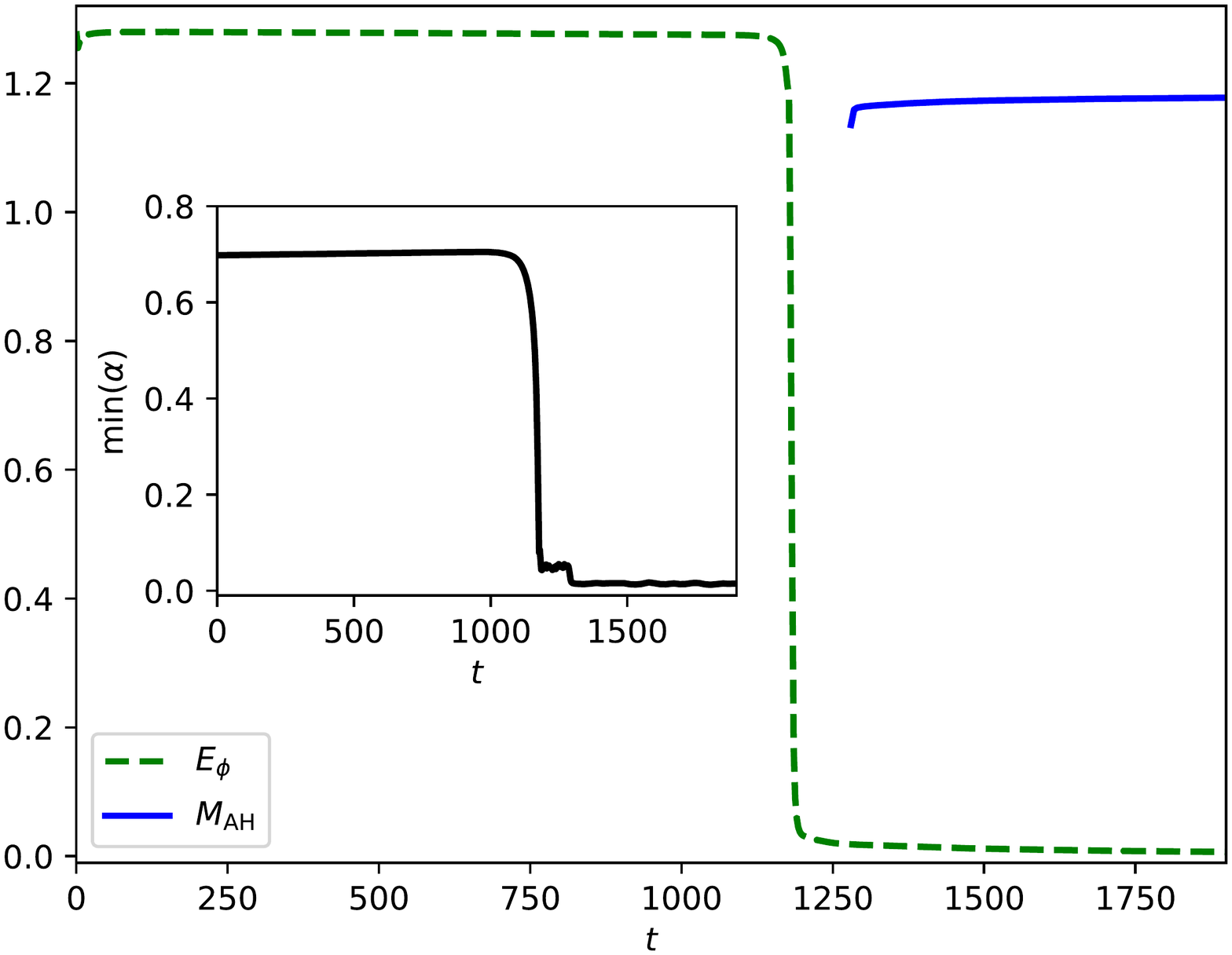}
\caption{Time evolution of a scalar SBS, model $2_S$. Six sequential snapshots of $\rho_E$ (top panels). Time runs left to right, $1^{st}$ to $2^{nd}$ row. Total scalar field energy and AH mass (main panel) and lapse function (inset).}
\label{fig3}
\end{figure}

First consider the scalar SBSs. Fig.~\ref{fig3} shows the time evolution of model $2_S$. Up to $t\sim 1000$ the star remains essentially undisturbed; then, following the development of a non-axisymmetric perturbation, 
see upper panels, the star pinches off into two fragments. 
The resulting binary is gravitationally bound and collapses into a BH at $t\sim 1200$. This is diagnosed by both the appearance of an apparent horizon (AH), whose mass is shown in the main panel, and the vanishing of the lapse function, $\alpha$, as seen in the inset of Fig.~\ref{fig3}.  A similar evolution is observed for model $1_S$.  This confirms that scalar SBSs, even in the FF, are prone to a non-axisymmetric instability. Unlike the formation scenario, here the instability leads to a complete gravitational collapse, likely due to the more compact initial data.

\begin{figure}[t!]
\begin{tabular}{ p{0.5\linewidth}  p{0.5\linewidth} }
\centering Model $1_P$ & \centering Model $3_p$
\end{tabular}
\\
\includegraphics[width=0.24\linewidth]{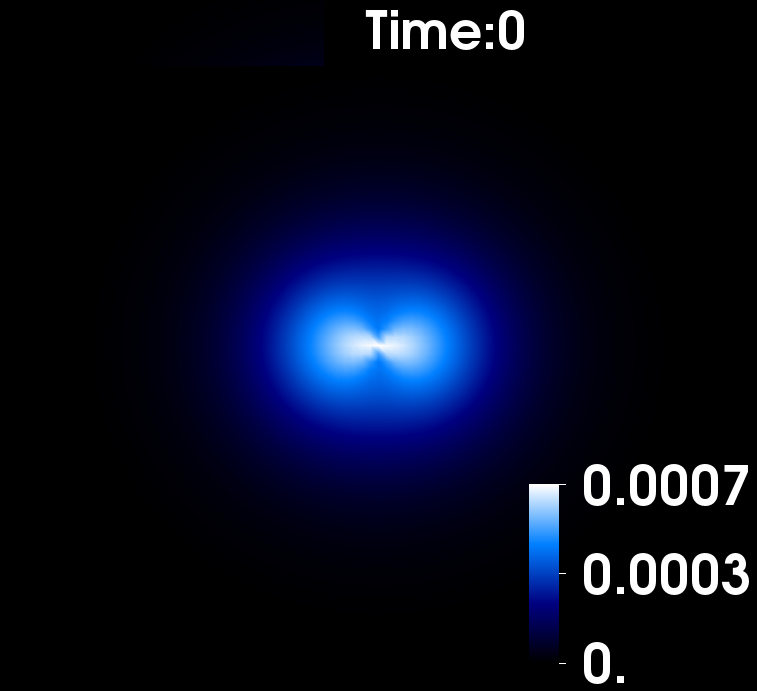}\hspace{-0.01\linewidth}
\includegraphics[width=0.24\linewidth]{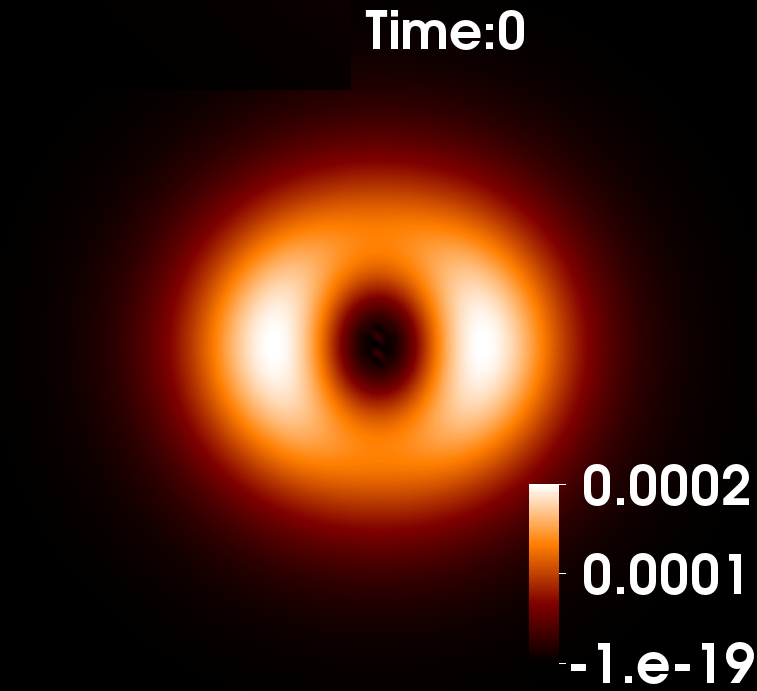}\hspace{0.02\linewidth}
\includegraphics[width=0.24\linewidth]{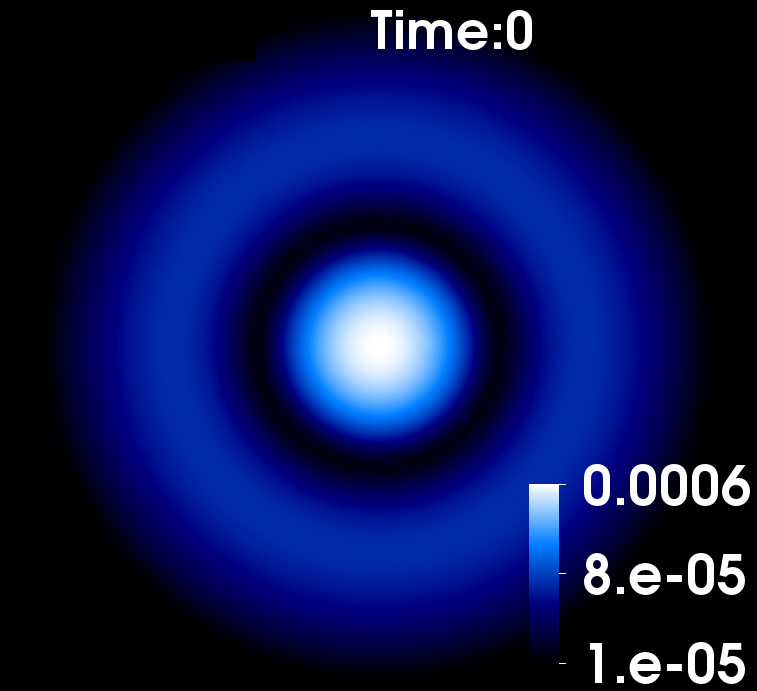}\hspace{-0.01\linewidth}
\includegraphics[width=0.24\linewidth]{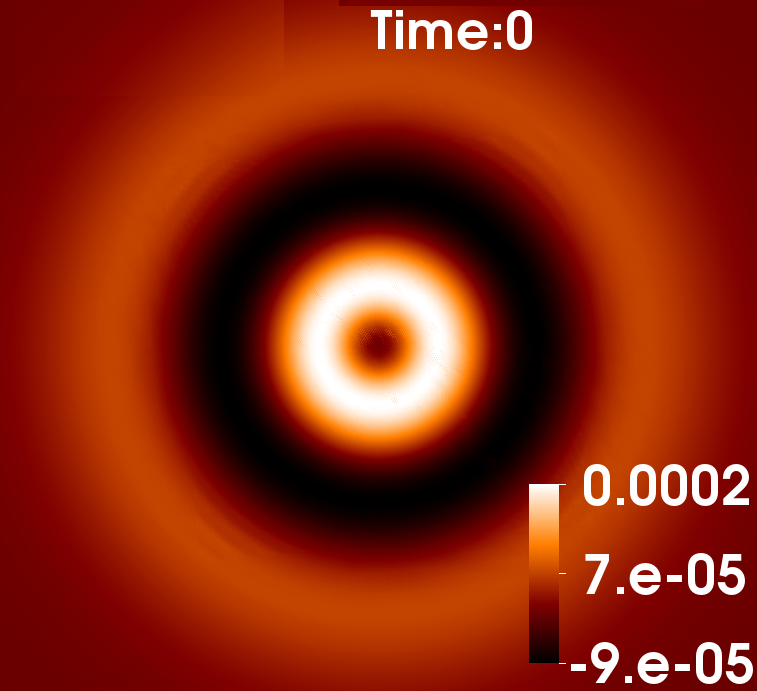}\\
\includegraphics[width=0.24\linewidth]{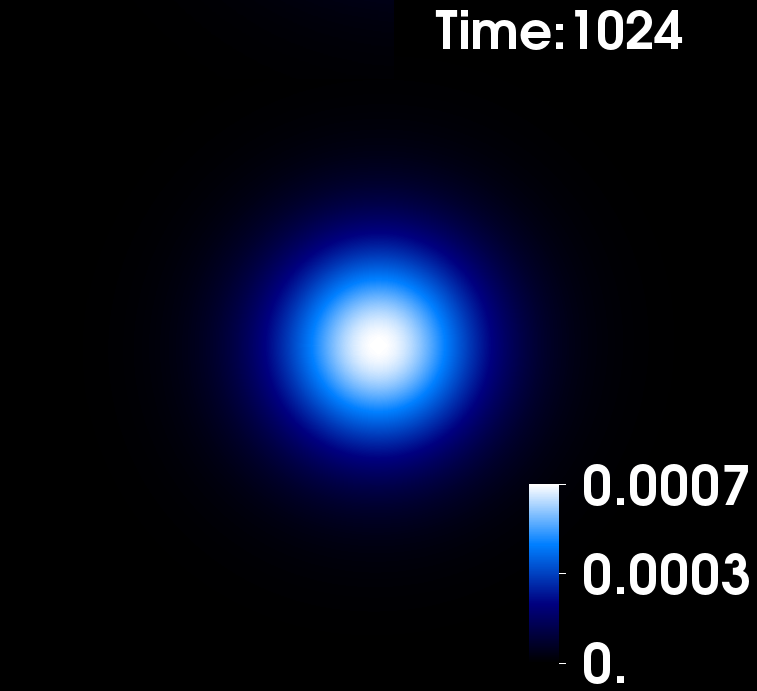}\hspace{-0.01\linewidth}
\includegraphics[width=0.24\linewidth]{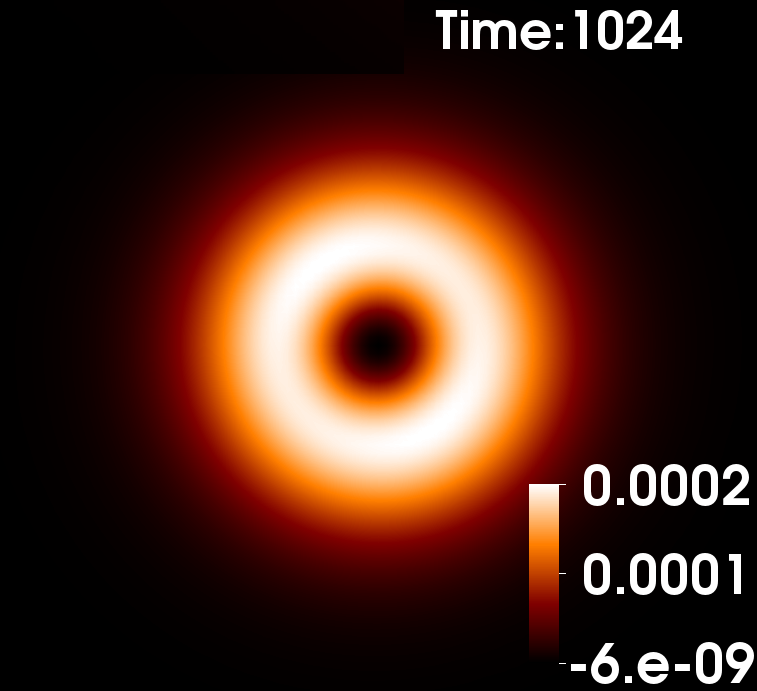}\hspace{0.02\linewidth}
\includegraphics[width=0.24\linewidth]{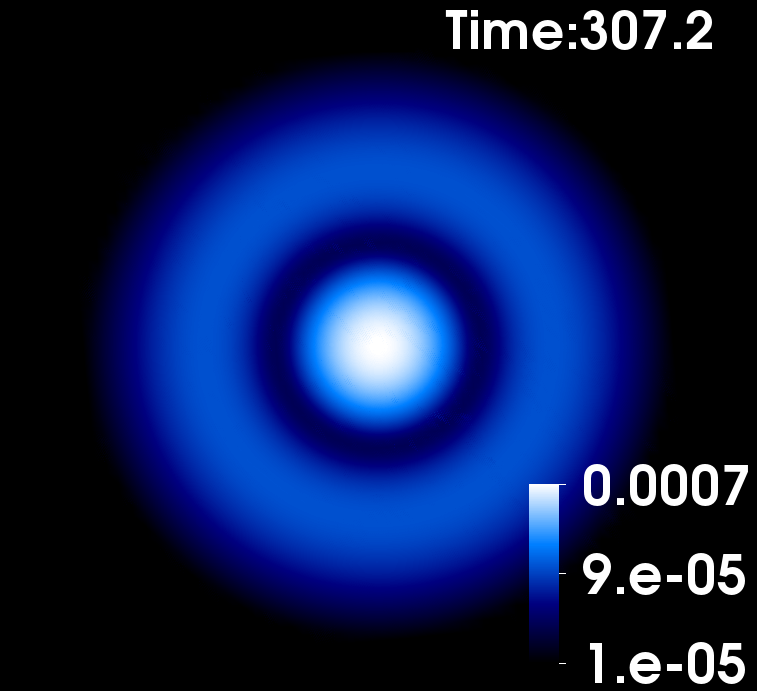}\hspace{-0.01\linewidth}
\includegraphics[width=0.24\linewidth]{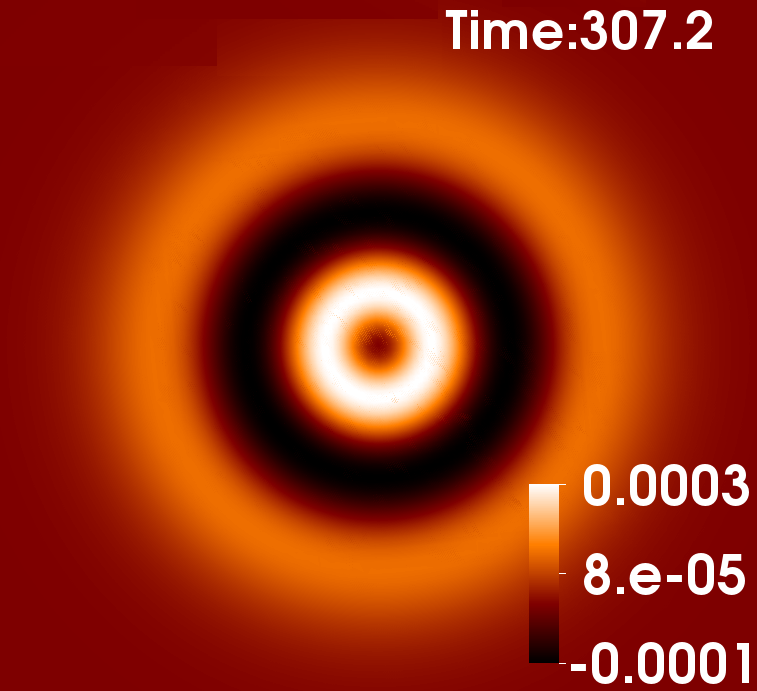}\\
\includegraphics[width=0.24\linewidth]{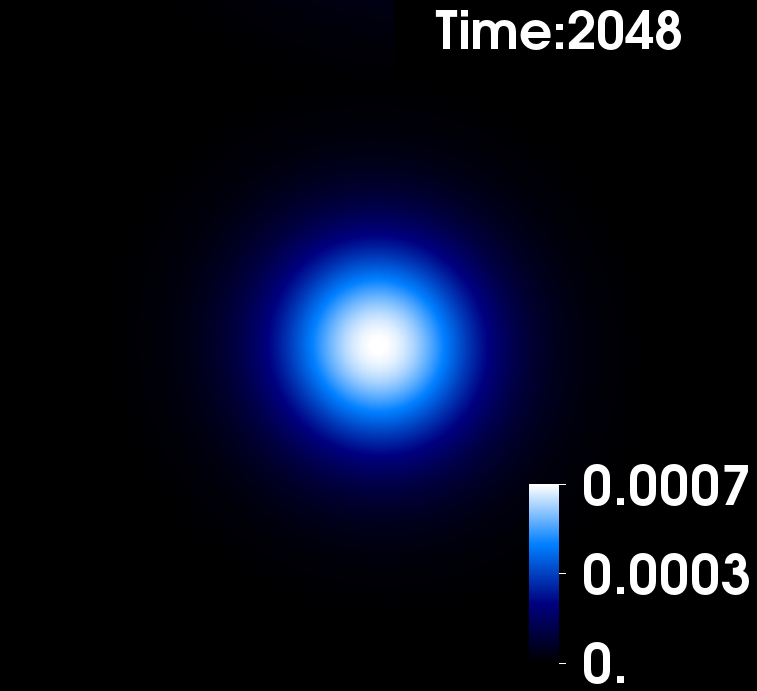}\hspace{-0.01\linewidth}
\includegraphics[width=0.24\linewidth]{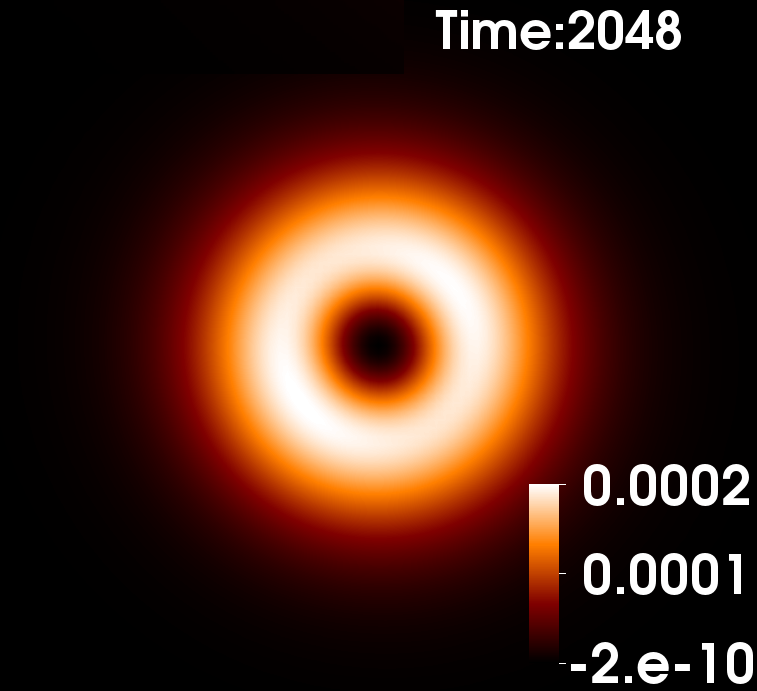}\hspace{0.02\linewidth}
\includegraphics[width=0.24\linewidth]{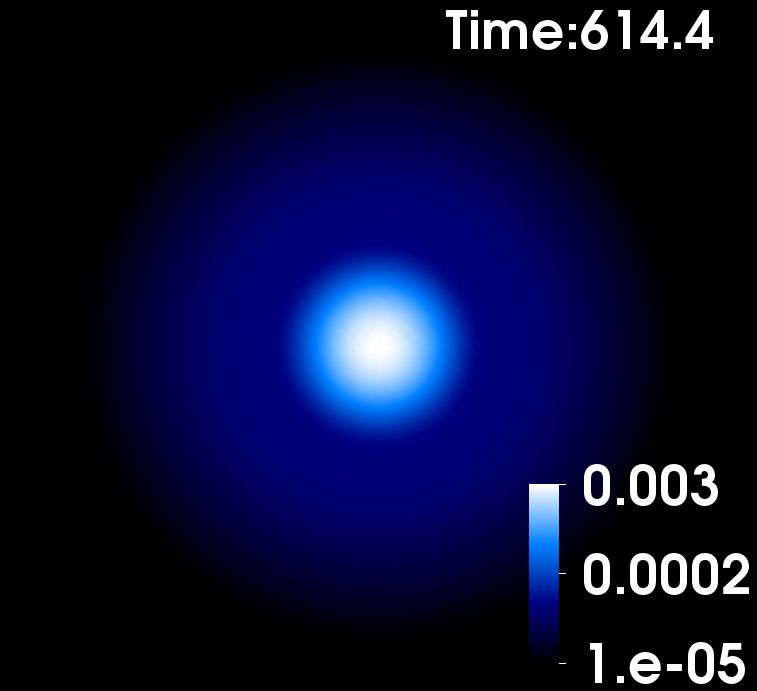}\hspace{-0.01\linewidth}
\includegraphics[width=0.24\linewidth]{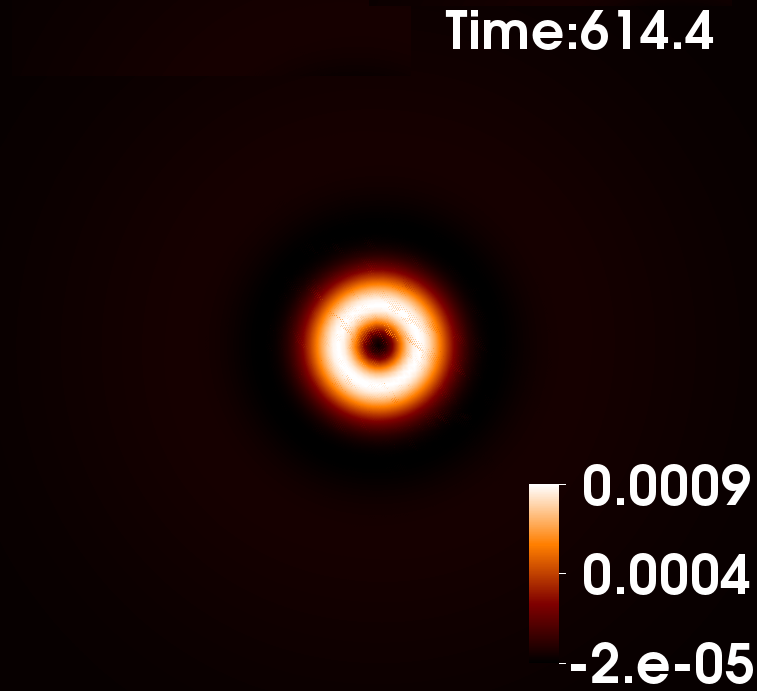}\\
\includegraphics[width=0.24\linewidth]{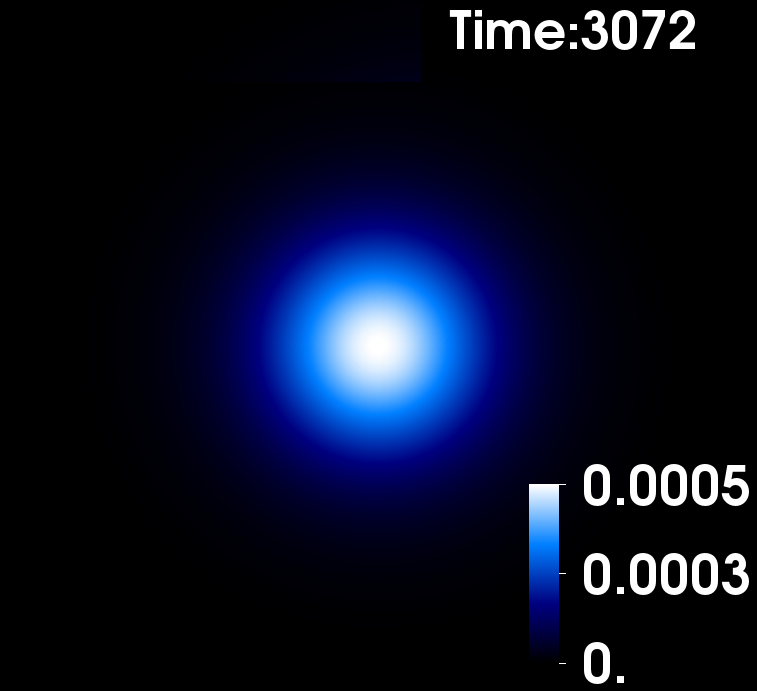}\hspace{-0.01\linewidth}
\includegraphics[width=0.24\linewidth]{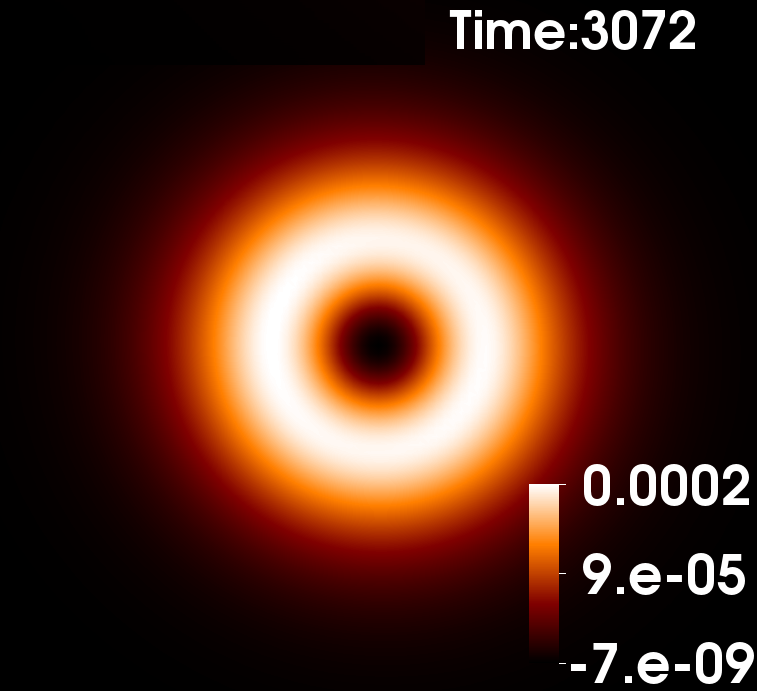}\hspace{0.02\linewidth}
\includegraphics[width=0.24\linewidth]{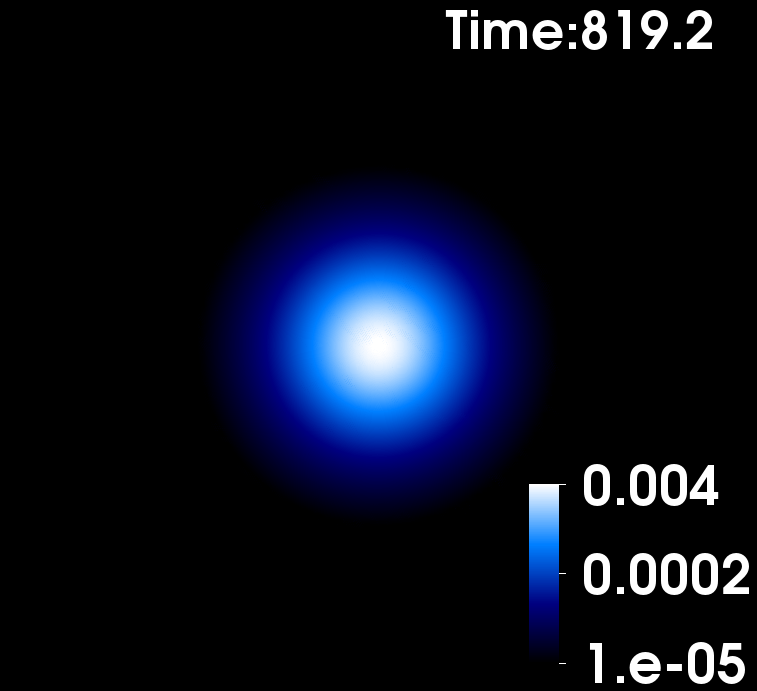}\hspace{-0.01\linewidth}
\includegraphics[width=0.24\linewidth]{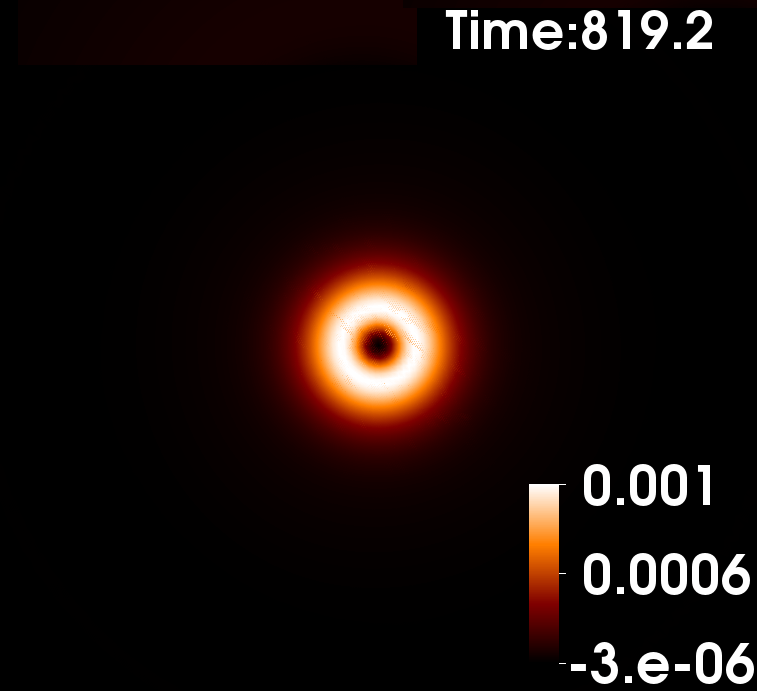}\\
\includegraphics[width=0.24\linewidth]{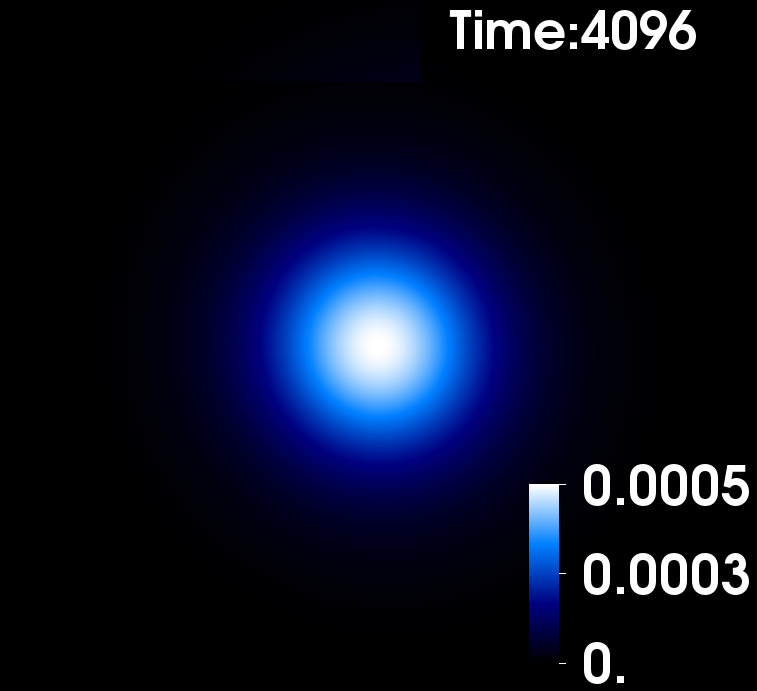}\hspace{-0.01\linewidth}
\includegraphics[width=0.24\linewidth]{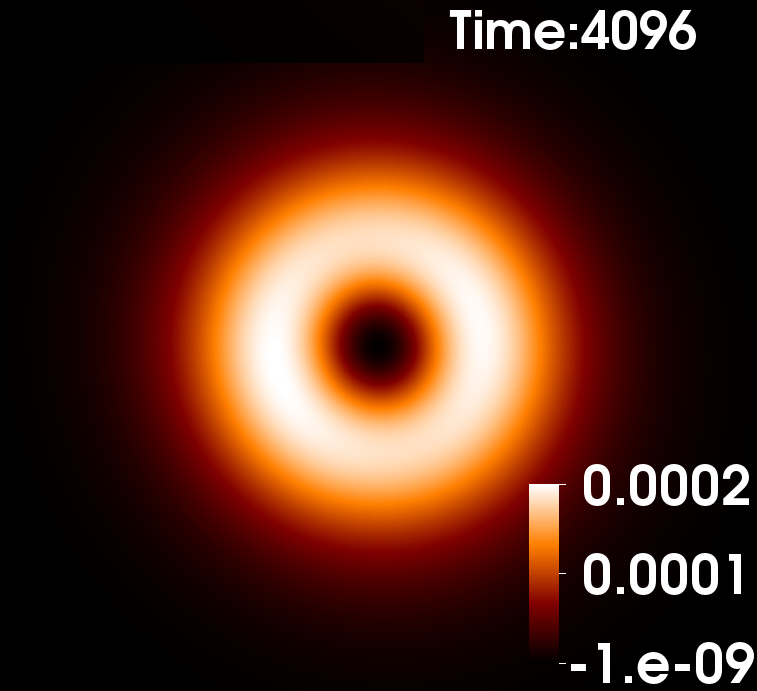}\hspace{0.02\linewidth}
\includegraphics[width=0.24\linewidth]{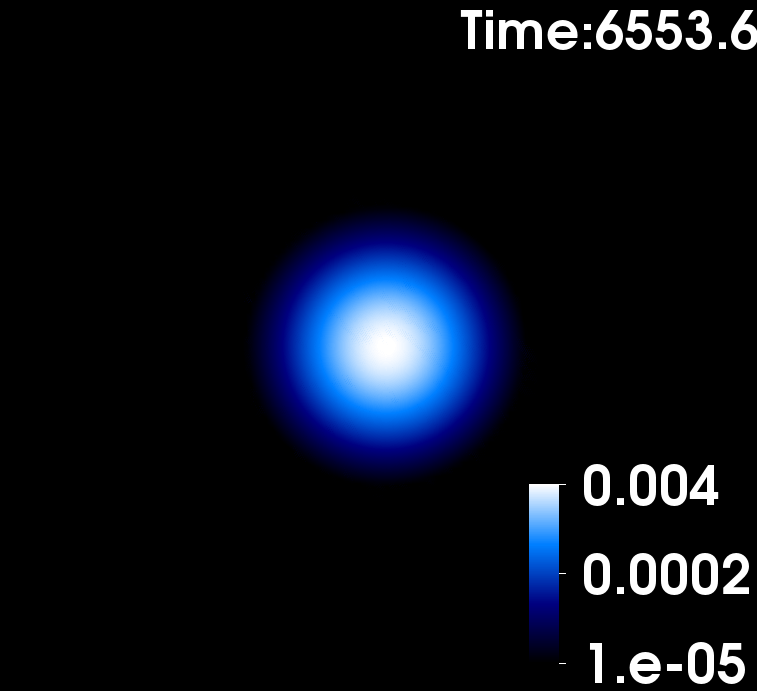}\hspace{-0.01\linewidth}
\includegraphics[width=0.24\linewidth]{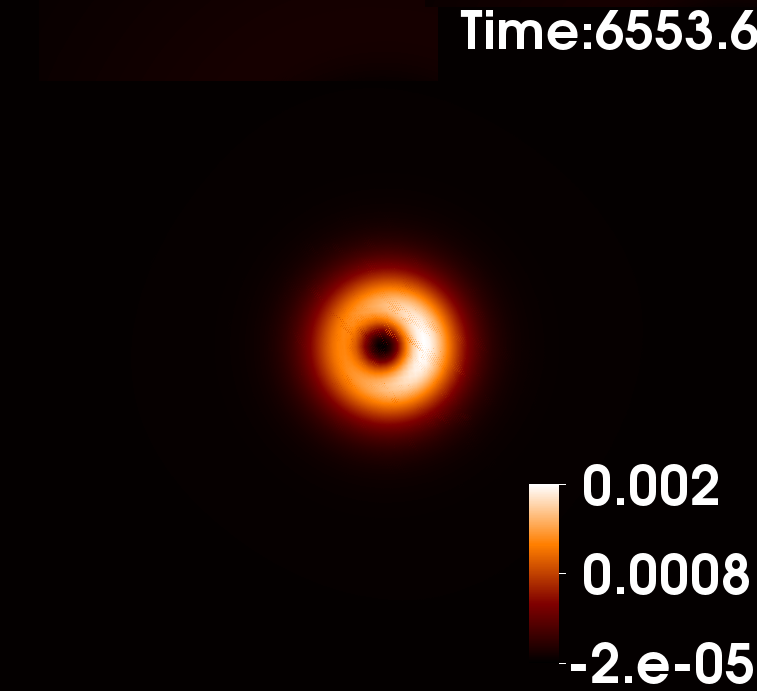}
\\
\caption{Time evolution of vector SBSs. Equatorial cut of $\rho_E$ (blue) and $\rho_J$ (orange) for the FF model $1_P$ with a perturbation  (left side) and the excited model $3_P$ (right side). 
}
\label{fig4}
\end{figure}

The behaviour of the vector SBSs is distinct.  FF solutions, such as models $1_P$ and $2_P$ show no sign of instability, in the absence of large perturbations. They neither disperse away nor collapse to a BH up to $t\sim 4000$, time at which the drift in the Proca field energy and angular momentum for  model $1_P$ is 2.0\% and 2.2\% respectively, whereas for model $2_P$, the drift is less than 1\%. We further tested the dynamical robustness of vector SBSs by perturbing models $1_P$ and $2_P$ and by considering some excited states, such as model $3_P$-$5_P$. Fig.~\ref{fig4} exhibits the time evolution of two examples: $(i)$ model $1_P$ with a perturbation of the sort considered in the formation scenario for the vector case, and with a sufficiently large amplitude to visibly distort the star (see first panel), and $(ii)$  the excited model $3_P$. In the first case, the perturbation, albeit large enough to deform the morphology of the star away from its spheroidal shape, is dissipated away, and the star recovers its shape. In the second case, the excited state structure of the star is manifest in the composite, Saturn-like, structure of its energy distribution~\cite{Herdeiro:2016tmi}. After $t\sim1000$, the star abruptly loses energy and angular momentum, until  $t\sim3000$ when it asymptotically tends to a new equilibrium configuration. This new configuration has no nodes and it is close to model $2_P$. Thus, the star migrates from the excited family to the FF, where it settles, advocating the stability of the latter. Excited models $4_P$ and $5_P$, on the other hand, collapse to a BH.

%

{\bf {\em Interpretation and further remarks.}} 
The contrasting dynamical properties of the scalar/vector SBSs break the phenomenological (qualitative) degeneracy observed between these two types of BSs in the spherical case. It is tempting to attribute this contrast to the different morphology of these stars, as exhibited in Fig.~\ref{fig0}. 
This interpretation is partly supported by the analogy with dynamical instabilities in differentially rotating relativistic (neutron) stars~\cite{Paschalidis:2016vmz}. In that case
the existence of a toroidal shape has been suggested to be a necessary condition for the development of non-axisymmetric corrotational instabilities~\cite{Saijo:2003,Watts:2005}.
In fact, the pinching instabilities and fragmentation exhibited in the scalar models above are reminiscent of evolutions of unstable toroidal fluid stars~\cite{Zink:2005rr,Zink:2006qa}, but also of other corrotational instabilities observed in toroidal systems such as  the Papaloizou-Pringle instability in accretions disks \cite{Kiuchi:2011}.
Preliminary results, moreover, indicate vector SBSs with $m=2$ (which are toroidal) are also unstable. The analogy with corrotational instabilities in relativistic fluid stars will be further explored elsewhere.

The instability of scalar SBSs may explain the inability to find them as endpoints in the evolution of orbiting binaries  of (non-spinning) BSs~\cite{Bezares:2017mzk,palenzuela2017gravitational}. By the same token, however, vector SBSs should form in the equivalent vector scenarios. This suggests revisiting the work in~\cite{sanchis2019head} using constraint-abiding initial data. A related question pertains the impact of matter self-interactions in the dynamics reported herein.


\bigskip

{\bf {\em Acknowledgements.}} 
We thank Carlos Palenzuela, Dar\'io N\'u\~nez, Juan Carlos Degollado, and Sergio Gimeno-Soler, for useful discussions and valuable
comments. This work has been supported by the Spanish MINECO (grant AYA2015-66899-C2-1-P), 
by the Generalitat Valenciana (ACIF/2015/216), by the Funda\c c\~ao para a Ci\^encia e a Tecnologia (FCT)
projects PTDC/FIS-OUT/28407/2017, UID/MAT/04106/2019 (CIDMA) and UID/FIS/00099/2013 (CENTRA), by national funds (OE), through FCT, I.P., in the scope of the framework contract foreseen in the numbers 4, 5 and 6
of the article 23, of the Decree-Law 57/2016, of August 29,
changed by Law 57/2017, of July 19. 
This work has further been supported by  the  European  Union's  Horizon  2020  research  and  innovation
(RISE) programmes H2020-MSCA-RISE-2015
Grant No.~StronGrHEP-690904 and H2020-MSCA-RISE-2017 Grant No.~FunFiCO-777740. We would like to acknowledge networking support by the COST Action GWverse
CA16104.
MZ acknowledges financial support provided by FCT/Portugal through the IF
programme, grant IF/00729/2015.
PC acknowledges the Ramon y Cajal funding (RYC-2015-19074) supporting his research.
Computations have been performed at the Servei d'Inform\`atica de la Universitat
de Val\`encia, on the ``Baltasar Sete-Sois'' cluster at IST, and at MareNostrum
(via PRACE Tier-0 Grant No. 2016163948).


\bigskip


{\bf {\em Appendix A. Initial data and numerical evolutions.}} 
To perform numerical relativity evolutions,  one must construct an initial configuration for the spacetime and matter fields that satisfies all the Einstein constraint equations. We solve numerically the Hamiltonian and momentum constraints using the extended conformally flatness condition (XCFC) approximation \cite{CorderoCarrion:2008nf}. For fast rotating neutron stars XCFC has been shown to provide accurate 
space-times within $1\%$ of full GR. In the following we sketch the main steps for our application.

In the $3+1$ spacetime decomposition the metric can be expressed as
\begin{equation}\label{4metric}
g_{\mu\nu}dx^{\mu}dx^{\nu} = -\alpha^2 dt^2 + \gamma_{ij} (dx^i + \beta^i dt) (dx^j + \beta^j dt) \ ,
\end{equation}
where $\alpha$ is the lapse function, and $\beta^i$ is the shift vector. We apply a conformal decomposition of the spatial 3-metric
\begin{equation}\label{metric}
\gamma_{ij} = \Psi^4 \widetilde{\gamma}_{ij} \ ,
\end{equation}
where $\Psi$ is the conformal factor. In XCFC, $\widetilde{\gamma}_{ij}$ is approximated by the flat 3-metric, that simplifies notably
the constraint equations. The procedure we follow to solve the constraints is the following. 
\begin{enumerate}
\item Make an ansatz for the matter fields (which we will detail below for the scalar and vector case).
\item Set the geometry, as a first step, to be Minkowski spacetime: $\alpha=\Psi=1$ and $\beta^i=0$. Then evaluate the matter source terms in the Hamiltonian and momentum constraints.
\item Having obtained the matter source terms, solve the constraint equations, obtaining an improved (non-Minkowski) value for $\alpha$, $\Psi$ and $\beta^i$.
\item With the updated values for the metric quantities, evaluate again the matter fields and matter source terms. Then, repeat the procedure iteratively.
\end{enumerate}
At each iteration, we evaluate the L2-norm of the difference between the updated and the previous step values. If it is lower than a certain tolerance ($10^{-8}$), 
we consider it has converged to a constraint-satisfying solution.

Let us now consider in detail the ansatz for both the scalar and Proca field, and the explicit form of the matter source terms.

\subsection{Scalar case} \label{section:scalar}
The evolution formulation adopted is the one described in \cite{Okawa:2014nda}, which involves the scalar field $\phi$ and the conjugated momenta $\Pi$ defined as 
\begin{equation}\label{Pi}
\Pi = -\frac{1}{\alpha} (\partial_t - \beta^i\partial_i)\,\phi \ .
\end{equation}
We specify the ``shape" of the scalar cloud as
\begin{equation}\label{Phi1}
 \phi(t, r, \theta, \varphi) = R(r) Y_{11}(\theta,\varphi)\,e^{-i\omega t} \ ,
\end{equation}
where $Y_{11}(\theta,\varphi)=\sin{\theta}\,e^{i\varphi}$ is the $\ell=m=1$ spherical harmonic and $R(r) = A\,r\, e^{-\frac{r^2}{\sigma^2}}$. At $t=0$, 
\begin{equation}\label{Pi2}
\Pi = -\frac{i}{\alpha} (\omega + \beta^{\varphi})\,\phi\ ,
\end{equation}
where we use that $\beta^{\varphi}$  is the only non-zero component of the shift vector,
a consequence of the axisymmetry invariance of the energy-momentum tensor. 
Note that the definition of $R(r)$ includes a factor $r$ that cancels out the $1/r^2$  factors  appearing in the matter source terms, due to the spherical spatial coordinates, and 
yields a well-defined behaviour in the limit $r\rightarrow 0$.

\subsection{Proca case} \label{section:Proca}
The evolution formulation adopted is the one described in \cite{Zilhao:2015tya}, which involves the scalar potential $\Aphi$, the vector potential $\A_{i}$ and the ``electric" and ``magnetic" fields $E^{i}$ and $B^{i}=\epsilon^{ijk}D_{j}\A_{k}$, where $D_{i}$ is the spatial covariant derivative and $\epsilon^{ijk}$ is the anti-symmetric Levi-Civita symbol.  In the Proca case, besides solving the Hamiltonian and momentum constraints, it is necessary to solve the Gauss constraint which reads
\begin{equation}\label{Gauss}
D_{i}E^i=-\mu^2 \Aphi \ .
\end{equation}
 We assume $E^i$ is conservative; thus it is the gradient of a potential $V$, $E^i= - \nabla^i V$. In this way, the Gauss constraint becomes
\begin{equation}\label{Gauss_V}
\nabla^2 V = \mu^2\Aphi \ .
\end{equation}
Now, the ansatz for the scalar potential is chosen to be 
\begin{equation} \label{scalar_potential}
\Aphi(t, r, \theta, \varphi)  = R(r) Y_{11}(\theta,\varphi)\,e^{-i\omega t} \ ,
\end{equation}
where $R(r) = A\,r^2 e^{-\frac{r^2}{\sigma^2}}$, and we solve equation \eqref{Gauss_V} to obtain $V$ and, thus, $E^i$. Due to the fact that the spherical harmonics $Y_{lm}(\theta,\varphi)$ are eigenfunctions of the operator $r^2\nabla^2$ with eigenvalues $-\ell(\ell+1)$, Eq.~\eqref{Gauss_V} can be reduced to an ordinary differential equation which depends only on the radial part and reads, at $t=0$,
\begin{equation} \label{Gauss2}
\frac{1}{r}\partial^2_r [r V(r)]  - \frac{2}{r^2} V(r) = \mu^2 R(r) \ .
\end{equation}

Similarly to the scalar case, the $r^2$ factor in $R(r)$  is needed to have a well-defined behaviour in the limit $r\rightarrow 0$. We solve Eq.~\eqref{Gauss2} analytically specifying boundary conditions $\lim_{r \to 0} V(r) = 0$ and $\lim_{r \to \infty} V(r) = 0$ to find
\begin{equation}
\begin{split}
V(r) =& \frac{A \,\sigma e^{-\frac{r^2}{\sigma^2}}}{24 r^2} \biggl( -e^{\frac{r^2}{\sigma^2}}\biggl[2 \sqrt{\pi} \sigma^2 r^3 + 8 \sigma^5  \\
		 & -2\sqrt{\pi}\sigma^3r^2 \,\rm{Erf} \left(\frac{r}{\sigma}\right)\biggr] + 8 r^2 \sigma^3 + 8 \sigma^5 \biggr) \ .
\end{split}
\end{equation}

From the potential $V$ we can evaluate $E^i$ but we still need to specify a shape for the vector potential $\A_r$. The XCFC formalism assumes maximal slicing, $K=0$, where $K$ is the trace of the extrinsic curvature. Therefore, the evolution equation for the scalar potential takes the form
\begin{equation}\label{phi_equation}
\partial_t \Aphi = -\A^i D_i\alpha - \alpha D_i \A^i + \beta^{\varphi}\partial_{\varphi}\Aphi \ ,
\end{equation}
where we have considered that the only non-zero component of the shift vector is $\beta^{\varphi}$, as in the scalar case. Using 
Eq.~\eqref{scalar_potential} we get
\begin{equation} \label{ai_equation}
D_i(\alpha \A^i) = i (\omega + \beta^{\varphi}) \Aphi \ ,
\end{equation}
which has the same form as \eqref{Gauss} and can be solved similarly, yielding
\begin{equation}
\A^i = \frac{i}{\alpha} (\omega + \beta^{\varphi}) E^i \ .
\end{equation}
Since $\A^i$ can be written as the gradient of a potential function, it is an irrotational field. Therefore the ``magnetic" field for these initial data is zero by construction. 

As a summary, the $t=0$ initial data for the fields read
\begin{align}
\Aphi(r,\theta,\varphi) =&\ A\,r^2\,e^{-\frac{r^2}{\sigma^2}} \sin{\theta} e^{ i\varphi} \ , \nonumber \\
E^r(r,\theta,\varphi) =&\ -\hat{E}^r(r,\theta) e^{ i\varphi} \ , \nonumber \\
E^{\theta}(r,\theta,\varphi) =&\ -\hat{E}^{\theta}(r,\theta) e^{ i\varphi} \ , \nonumber  \\
E^{\varphi}(r,\theta,\varphi) =&\ -\hat{E}^{\varphi}(r,\theta) e^{ i(\varphi+\pi/2)}\ , \nonumber \\
\A_r(r,\theta,\varphi) =&\ \hat{\A}_r(r,\theta)e^{ i(\varphi+\pi/2)} \ , \nonumber \\
\A_{\theta}(r,\theta,\varphi) =&\ \hat{\A}_{\theta}(r,\theta)e^{ i(\varphi+\pi/2)}\ , \nonumber \\
\A_{\varphi}(r,\theta,\varphi) =&\ -\hat{\A}_{\varphi}(r,\theta)e^{ i\varphi} \ , 
\end{align}
where the $r$ and $\theta$ dependence is: 
\begin{align}
\hat{E}^r(r,\theta) =&\ \frac{A\, \sigma^2}{12 r^3} \biggl( 8 \sigma^4 - 2 e^{-\frac{r^2}{\sigma^2}} [3 r^4 +  \\
     &\ 4r^2\sigma^2 + 4\sigma^4] -\sqrt{\pi}r^3\sigma \,\rm{Erf} \left(\frac{r}{\sigma}\right) \biggr)\sin{\theta}\ , \notag \\
\hat{E}^{\theta}(r,\theta) =&\ \frac{A\, \sigma e^{-\frac{r^2}{\sigma^2}}}{24 r^3} \biggl( -e^{\frac{r^2}{\sigma^2}}\biggl[2 \sqrt{\pi} \sigma^2 r^3 + 8 \sigma^5  \\
		 & -2\sqrt{\pi}\sigma^3r^2 \,\rm{Erf} \left(\frac{r}{\sigma}\right)\biggr] + 8 r^2 \sigma^3 + 8 \sigma^5 \biggr) \cos{\theta}\ , \notag \\
\hat{E}^{\varphi}(r,\theta) =&\ \frac{A\, \sigma e^{-\frac{r^2}{\sigma^2}}}{24 r^3} \biggl( -e^{\frac{r^2}{\sigma^2}}\biggl[2 \sqrt{\pi} \sigma^2 r^3 + 8 \sigma^5  \\
		 & -2\sqrt{\pi}\sigma^3r^2 \,\rm{Erf} \left(\frac{r}{\sigma}\right)\biggr] + 8 r^2 \sigma^3 + 8 \sigma^5 \biggr) \cos{\theta}\ , \notag \\
\hat{\A}_i(r,\theta) =&\ \frac{1}{\alpha} (\omega + \beta^{\varphi}) \gamma_{ij}\hat{E}^j(r,\theta) \ .
\end{align}

{\bf {\em Appendix B. Code Assessment.}} 
For the numerical evolutions we employ the codes introduced in
Refs.~\cite{Cunha:2017wao,sanchis2019head}, which make use of the \textsc{EinsteinToolkit}
infrastructure~\cite{Loffler:2011ay,EinsteinToolkit:web,Zilhao:2013hia} with the
\textsc{Carpet} package~\cite{Schnetter:2003rb,CarpetCode:web} for
mesh-refinement capabilities and
\textsc{AHFinderDirect}~\cite{Thornburg:2003sf,Thornburg:1995cp} for finding
apparent horizons.
The evolution of the spacetime metric is handled either by \textsc{Lean}~\cite{ZilhaoWitekCanudaRepository} - originally
presented in~\cite{Sperhake:2006cy} for vacuum spacetimes, and now also
distributed within the \textsc{EinsteinToolkit} infrastructure - or \textsc{McLachlan}~\cite{Brown:2008sb}.

\begin{figure}[h!]
\centering
\includegraphics[height=2.45in]{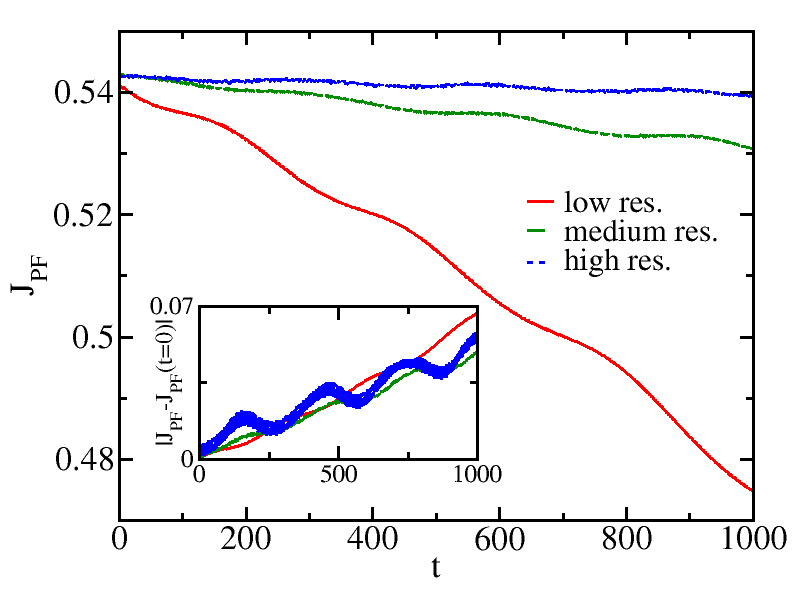}
\caption{Time evolution of the angular momentum of model $1_P$ for three different resolutions. (Inset) The L1-norm rescaled to second order convergence.}
\label{fig11}
\end{figure}

The numerical code we have used to evolve the rotating Proca stars has been assessed in~\cite{Zilhao:2015tya} for a real Proca field and in~\cite{sanchis2019head} for the complex case. In this section we will discuss the convergence analysis of the rotating Proca star stable solution. We performed the evolution of model $1_P$ with three different resolutions, namely with high resolution, corresponding to $\lbrace(96, 48, 24, 12)$, $(1.6, 0.8, 0.4, 0.2)\rbrace$, medium resolution grid structure of $\lbrace(96, 48, 24, 12)$, $(3.2, 1.6, 0.8, 0.4)\rbrace$ and low resolution $\lbrace(96, 48, 24, 12)$, $(6.4, 3.2, 1.6, 0.8)\rbrace$. The first set of numbers indicates the spatial domain of each level and the second set indicates the resolution.
In Fig.~\ref{fig11} we show the time evolution of the angular momentum for the different resolutions. The lower the resolution, the larger is the drift at $t=1000$: 12.3\% for low resolution, 2.3\% for medium resolution and 0.6\% for the high resolution. The L1-norm can be rescaled to second order convergence as shown in the inset of Fig~\ref{fig11}.
For further details concerning the performance of the codes employed, please
see~\cite{Cunha:2017wao,ZilhaoWitekCanudaRepository} for the scalar code and~\cite{sanchis2019head} for the
Proca one.

\bibliography{num-rel2}

\end{document}